%% file: feast.tex
\documentclass{article} 
\usepackage{latexsym,saad}
\usepackage{amssymb,amsmath,amsfonts,url,hyperref} 

\usepackage[pdftex]{graphicx} 
\usepackage[ruled,vlined,linesnumbered]{algorithm2e}
\usepackage{caption}
\usepackage{verbatim} 
\usepackage{subcaption}
\usepackage{color}          
\usepackage{cite}
\usepackage{framed}

\usepackage[table]{xcolor}
\usepackage{fancyhdr}
   
\newcommand{\col}{\cellcolor{gray!15}}

\usepackage{wrapfig,lipsum} 

\usepackage{multirow}
\graphicspath{{./FIGS/}} 
\usepackage{graphicx,epsf,picinpar}
\usepackage{listings}
\newcommand{\revd}[1]{{\color{red} #1}}

\newsavebox{\fmbox}

\newcommand{\versionT}{{3.0}}
\newcommand{\version}{{v\versionT}{\ }}
\newcommand{\feastdir}{{\tt <}{\it {\tt FEAST} directory}{\tt >~}}
\newcommand{\arch}{{\tt <{\it arch}>~} }

\newcommand{\BS}{{$\backslash$}}

\newcommand{\Indent}{\hspace*{0.2in}}

\begin{document}

\pagestyle{empty}

\centerline{\sc High-Performance Numerical Library for Solving Eigenvalue Problems}

\vspace{4cm}
\begin{center}
{\Huge FEAST Eigenvalue Solver \version\\[10pt]
User Guide \\[15pt] {\Large Eric Polizzi, James Kestyn}} \\ 
\vspace{2.2cm}
\scalebox{0.7}{\includegraphics{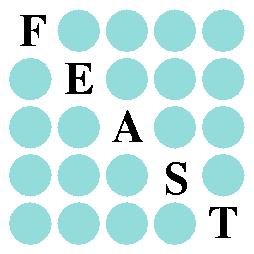}}\\
{\tt http:://www.feast-solver.org}
\end{center}

\vspace{3.5cm}
\begin{center}
\begin{Large}
Eric Polizzi Research Lab. \\
Department of Electrical and Computer Engineering, \\
Department of Mathematics and Statistics, \\[5pt]
University of Massachusetts, Amherst
\end{Large}
\end{center}

\newpage

\section*{References}

\noindent If you are using FEAST, please consider citing one or more publications below in your work.

\begin{small}
\noindent \begin{description}
\item[{\bf Main reference}]~ \\ E. Polizzi, {\it Density-Matrix-Based Algorithms for Solving Eigenvalue Problems},\\
 Phys. Rev. B. Vol. 79, 115112 (2009)
\item[{\bf Math analysis}]~ \\ P. Tang, E. Polizzi, {\it FEAST as a Subspace Iteration EigenSolver Accelerated by Approximate Spectral Projection};
SIAM Journal on Matrix Analysis and Applications (SIMAX)  35(2), 354–390 - (2014) 

\item[{\bf Non-Hermitian solver}]~ \\ J. Kestyn, E. Polizzi, P. T. P. Tang, 
{\it FEAST Eigensolver for Non-Hermitian Problems}, \\ 
arxiv.org/abs/1506.04463 (2015)

\item[{\bf Hermitian using Zolotarev quadrature}]~ \\ S. G\"uttel, E. Polizzi, P. T. P. Tang, G. Viaud, {\it
Optimized Quadrature Rules and Load Balancing for the FEAST Eigenvalue Solver}, \\ 
SIAM Journal on Scientific Computing (SISC), to appear (2015), arxiv.org/abs/1407.8078 (2014)

\item[{\bf Eigenvalue count using stochastic estimates}]~ \\ 
E. Di Napoli, E. Polizzi, Y. Saad, {\it
Efficient Estimation of Eigenvalue Counts in an Interval},\\ 
arxiv.org/abs/1308.4275 (2015)


\end{description}
\end{small}


\section*{Contact}

If you have any questions or feedback regarding FEAST, please send an-email to  
{\bf feastsolver@gmail.com}.

\vspace{1.0cm}

\section*{FEAST algorithm and software team, collaborators and contributors}

\begin{small}
\begin{center}
\begin{tabular}{|c|lp{5cm}p{5cm}|}
 \multicolumn{2}{c}{} & \multicolumn{1}{l}{\bf Algorithm and research} & \multicolumn{1}{l}{\bf Software development} 
 \\[5pt] \hline
\col & {\bf Eric Polizzi} &   Lead \newline &  v3.0, v2.1, v2.0, v1.0 \\ \cline{2-4}
\col &{\bf James Kestyn} &  Non-Hermitian FEAST  
  & v3.0  \\
\col {\bf UMass} &                  &      improved schemes, tunings, tools            &  FEAST non-Hermitian upgrade\\  \cline{2-4}
\col {\bf Amherst}  & {\bf Brendan Gavin} &   Non-linear eigenvector problem & \\ 
\col \col {\bf Team} & &   expanded subspace scheme &  \\  \cline{2-4}
\col & {\bf Braegan Spring} &  SPIKE-SMP v1.0  & v3.0  \\
\col &   & scalable banded system solver  & FEAST banded interfaces upgrade \\  \hline\hline
\col & {\bf Peter Tang} &  General FEAST algorithm analysis   &  \\  
\col  & Intel &    improved schemes  &  \\  \cline{2-4} 
\col {\bf Collaborators}& {\bf Yousef Saad} & Eigenvalue count estimates &   \\
\col {\bf \&} & U. of Minnesota & new  FEAST schemes (in progress) &   \\ \cline{2-4}
\col {\bf Contributors} & {\bf Edoardo Di Napoli}  & Eigenvalue count estimates  &     \\
\col & J\"ulich Supercomput.&  & \\ \cline{2-4}
\col & {\bf Stefan G\"uttel} &  Zolotarev quadrature \& analysis & v3.0  \\ 
\col  & Manchester U.         &  & Zolotarev quadrature database  \\ \cline{2-4}
\col & {\bf Gautier Viaud}  &  Zolotarev quadrature \& analysis  &  \\ 
\col &  ECP France&  & \\  \cline{2-4}
\col  & {\bf Ahmed Sameh} &  SPIKE-SMP v1.0 & \\ 
\col & Purdue U.&   SVD-FEAST (in progress) &    \\ \hline
\end{tabular}
\end{center}
\end{small}

\vspace{1cm}

\section*{Acknowledgments}

We acknowledge the many 
 helpful technical discussions and inputs from Dr. Sergey Kuznetsov and team members from Intel-MKL.
This work has been partially supported by Intel Corporation. 

\newpage

\tableofcontents

\newpage

\input{feast_overview}
\input{feast_interfaces}

\input{feast-use}

\input{feast_applications}

\input{feast_tools}




\end{document}

%% file: feast_overview.tex
\section{Updates/Upgrades Summary}

\begin{center}{\em If you are a FEAST's first time user, you can skip this section.}\end{center}
 Here is a summary of the most important updates/upgrades.

\subsection{From v2.1 to v3.0}
\begin{itemize}
\item A variety of new features have been added in v3.0. This includes support for non-Hermitian matrices, elliptical contours 
and custom user-defined contours, stochastic estimates for the number of eigenvalues inside search interval, 
and different quadrature rules. Many new routines have been added. See Table \ref{v3newroutines} for a summary.
\end{itemize}

\begin{table}[htbp]
\begin{small}
\begin{center}
\begin{tabular}{|l||ll|} \hline 
       Family of Eigenvalue Problems & \multicolumn{2}{c|}{Routines in v3.0} \\
$AX=BX\Lambda$ & & \\
$A^H\widehat{X}=B^H\widehat{X}\Lambda^*$ & Elliptical Contours (Standard) & Custom Contour (Expert)\\ \hline \hline
       Real and Symmetric & \footnotesize \tt \{s,d\}feast\_srci & \footnotesize \revd{\tt \{s,d\}feast\_srcix}  \\
       $A=A^T$, $B$ spd, $X=\widehat{X}$ real, $\Lambda$ real & \footnotesize \tt\{s,d\}feast\_\{sy,sb,scsr\}\{ev,gv\}  & \footnotesize \revd{\tt\{s,d\}feast\_\{sy,sb,scsr\}\{ev,gv\}x}  \\ \hline
       Complex and Hermitian &\footnotesize \tt \{c,z\}feast\_hrci & \footnotesize \revd{\tt \{c,z\}feast\_hrcix}  \\
$A=A^H$, $B$ hpd, $X=\widehat{X}$ complex, $\Lambda$ real                            & \footnotesize \tt \{c,z\}feast\_\{he,hb,hcsr\}\{ev,gv\}  & \footnotesize \revd{\tt \{c,z\}feast\_\{he,hb,hcsr\}\{ev,gv\}x}  \\ \hline
 Complex and Symmetric & \footnotesize \revd{\tt \{c,z\}feast\_srci} & \footnotesize \revd{\tt \{c,z\}feast\_srcix} \\
 $A=A^T$, $B=B^T$, $X=\widehat{X}^*$ complex, $\Lambda$ complex                             & \footnotesize \revd{\tt \{c,z\}feast\_\{sy,sb,scsr\}\{ev,gv\}} & \footnotesize \revd{\tt \{c,z\}feast\_\{sy,sb,scsr\}\{ev,gv\}x} \\ \hline
       Real and Non-Symmetric & \footnotesize \revd{\tt \{s,d\}feast\_grci} & \footnotesize \revd{\tt \{s,d\}feast\_grcix} \\
$A,B$ general, $X\neq\widehat{X}$ complex, $\Lambda$ complex                              & \footnotesize \revd{\tt \{s,d\}feast\_\{ge,gb,gcsr\}\{ev,gv\}} & \footnotesize \revd{\tt \{s,d\}feast\_\{ge,gb,gcsr\}\{ev,gv\}x} \\ \hline
       Complex and General & \footnotesize \revd{\tt \{c,z\}feast\_grci} & \footnotesize \revd{\tt \{c,z\}feast\_grcix} \\
$A,B$ general, $X\neq\widehat{X}$ complex, $\Lambda$ complex                              & \footnotesize \revd{\tt \{c,z\}feast\_\{ge,gb,gcsr\}\{ev,gv\}} & \footnotesize \revd{\tt \{c,z\}feast\_\{ge,gb,gcsr\}\{ev,gv\}x} \\ \hline
\end{tabular}
\caption{\label{v3newroutines} Summary of all routines in FEAST v3.0 (140 total) - new routines in red}
\end{center}
\end{small}
\end{table}

\begin{itemize}

\item Non-Hermitian routines use a different variant of the FEAST algorithm than Hermitian cases. 
The major difference is the use of dual subspaces, $Q$ and $\widehat{Q}$, corresponding to Right $X$ and Left $\widehat{X}$ eigenvectors. 
Also, the search interval must become 2-dimensional to account for complex eigenvalues. More detail is given in:

\noindent
 {\it FEAST Eigensolver for Non-Hermitian Problems,}\\ J. Kestyn, E. Polizzi, P. Tang, 
\url{http://arxiv.org/abs/1506.04463} (2015)

%

\item FEAST now offers multiple quadrature rules: Gauss, Trapezoidal and Zolotarev (for the Hermitian case), as well
as elliptical complex contour. 
More detail is given in:

\noindent
{\it Optimized Quadrature Rules and Load Balancing for the FEAST Eigenvalue Solver,} \\
S. G{\"u}ttel, E. Polizzi, P. T. Tang, G. Viaud, \url{http://arxiv.org/abs/1407.8078}

\item All FEAST routines can be called within their ``expert mode'' version which features new user input lists for 
nodes and weights.

\item Stochastic estimates for the number of eigenvalues inside of the search interval are now available. This feature can help
 users in estimating a value for the search subspace $M_0$. Refer to the following publication for more information.  

\noindent
{\it Efficient Estimation of Eigenvalue Counts in an Interval,} \\E. Di Napoli , E. Polizzi, Y. Saad,
\url{http://arxiv.org/abs/1308.4275}

\item Various utility routines have also been added (see section 6). We note in particular the possibility for the users
to design their own contour shape in the complex plane.  This is particularly helpful 
for non-Hermitian routines as it grants flexibility in targeting specific eigenvalues. See Section 6 for additional information.

\item FEAST PARAMETERS- new or updated {\tt fpm}  parameters:
\begin{itemize}

\item {\tt fpm(2)} is updated- inludes more options for \#nodes in the half-contour (for Hermitian FEAST) \\
If {\tt fpm(16)}=0,2, values permitted $[1 \mbox{ to } 20, 24, 32, 40 , 48, 56 ]$ \\
If {\tt fpm(16)}=1, all values permitted

\item {\tt fpm(6)} is updated- default value changed to 1 -\\
 Convergence criteria on trace (0) or eigenvectors relative residual (1)

\item {\tt fpm(8)} is added - Total number of contour integration nodes (i.e. complex shifts) for non-Hermitian FEAST. \\
If {\tt fpm(17)}=0, values permitted $[2 \mbox{ to } 40, 48, 64, 80 , 96, 112 ]$ \\
If {\tt fpm(17)}=1, all values permitted \\
 {\em Remark:}  {\tt fpm(8)} represents the \#nodes for the full contour while  {\tt fpm(2)} represents the 
\#nodes for the half-contour used by Hermitian FEAST.

\item {\tt fpm(10)} is added - can be used with the FEAST predefined driver interfaces (0: default, 1: store all the linear
 system factorizations).\\
             {\em Remark:} (i) storing the factorizations will significantly improve the performances (for FEAST DENSE in particular), 
but can significantly increase the memory usage;  
(ii) option 1 works with FEAST-MPI as well - store all factors associated to a given mpi process  

\item {\tt fpm(14)} is modified - include option 2  \\
 0- default normal FEAST execution  \\
 1- return only subspace Q size M0 after 1 contour         \\ 
  2- return stochastic estimates of the \#eigenvalue (in argument 'M' and 'res' for running average)

\item {\tt fpm(16)} is added -  Integration type for symmetric (0: Gauss/Default, 1: Trapezoidal, 2: Zolotarev)
\item {\tt fpm(17)} is added -  Integration type for non-symmetric (0: Gauss, 1: Trapezoidal/Default)
\item {\tt fpm(18)} is added -  Ratio for ellipsoid contour - fpm(18)/100 is ratio 'vertical axis'/'horizontal axis'
of the ellipse using the definition of the search contour. For example:\\
value 100 is the default (circle); \\ 
value 50 will create a 50\% flat ellipse; \\ 
value 200 will create a 200\% tall ellipse.  
\item {\tt fpm(19)} is added -  Rotation angle in degree [-180:180] for Ellipsoid contour and using FEAST non-Hermitian- Origin 
of the rotation is the vertical axis.
\end{itemize}

\end{itemize}

\newpage


\section{Preliminary}

{\small {\em 
``The solution of the algebraic eigenvalue problem has for long had a
particular fascination for me because it illustrates so well the difference
between what might be termed classical mathematics and practical
numerical analysis. The eigenvalue problem has a deceptively simple
formulation and the background theory has been known for many
years; yet the determination of accurate solutions presents a wide
variety of challenging problems.''}\\
J. H. Wilkinson- The Algebraic Eigenvalue Problem- 1965}

\vspace{0.3cm}

In his seminal textbook, J. H. Wilkinson artfully outlined the fundamentals, difficulties and numerical challenges for
addressing the eigenvalue problem. Since then, the eigenvalue problem has led to many challenging numerical questions and a
central problem: how can we compute eigenvalues and eigenvectors in an efficient manner and how
accurate are they? 

In many modern science and engineering applications, especially for those where the underlying system matrices are large and sparse, 
it is often the case that only selected segments of the eigenvalue spectrum are of interest. Although extensive efforts
have been devoted to develop new numerical algorithms and library packages, they are all commonly facing
new challenges for addressing the current large-scale simulations needs for ever higher level of accuracy,
robustness and scalability on modern parallel architectures. 
The FEAST eigensolver library package is intended to uniquely address all those issues. Its 
 originality lies with a new transformative numerical approach to the
traditional eigenvalue algorithm design - the FEAST algorithm.

\subsection{The FEAST Algorithm}

Unlike any other eigenvalue numerical software, the FEAST solver is based on a new  algorithm
 which deviates fundamentally from the  Krylov subspace based techniques 
(Arnoldi and Lanczos algorithms), Davidson-Jacobi techniques or other traditional subspace iteration techniques.
The FEAST algorithm is a general purpose eigenvalue solver
 which takes its inspiration from the density-matrix
 representation and contour integration technique in quantum mechanics\footnote{E. Polizzi, Phys. Rev. B. Vol.  79, 115112 (2009)}.
FEAST can be used for solving  the generalized eigenvalue problem \mbox{${AX}={BX}\Lambda$} (Hermitian or non-Hermitian), 
and obtaining all the eigenvalues $\lambda$ and (left/right) eigenvectors within a given 
search interval  $[\lambda_{min},\lambda_{max}]$ or 
an arbitrary contour in the complex plane.
FEAST's main building block is a numerical quadrature computation i.e. ${Q}=\sum w_j {Q_j}$, consisting of solving independent
linear systems  along a complex contour i.e. $(z_j{B}-{A}){Q_j}={Y}$ (with $z_j$ quadrature node), each with
multiple right hand sides ${Y}$. A Rayleigh-Ritz procedure is then used to generate
 a reduced dense eigenvalue problem orders of magnitude smaller 
than the original one
 (the size of this reduced problem is of the order of the
 number of eigenpairs inside the search interval/contour).
The algorithm contains
elements from complex analysis, numerical linear algebra and approximation theory, to produce
 an optimal subspace iteration method using spectral projectors\footnote{P. Tang, E. Polizzi, SIMAX  35(2), 354–390 - (2014)}.
Not only the FEAST algorithm features some unique and remarkable convergence and
robustness properties, it can exploit a key strength of modern computer architectures,
namely, multiple levels of parallelism.
All the important intrinsic properties of the algorithm, which have been
analyzed and commented at length in publications, can be summarized as follows:
\begin{itemize}
\itemsep 1pt
\parskip 1pt
\item[(i)] all multiplicities are naturally captured; 
\item[(ii)]
 no explicit orthogonalization procedure on long vectors is required;
\item[(iii)] reusable subspace capable to generate 
suitable initial guess;
\item[(iv)] allows the use of iterative methods for solving large-sparse linear systems; 
\item[(v)] can exploit 
natural parallelism at three different levels: \begin{enumerate}
\itemsep 1pt
\parskip 1pt 
\item search intervals can be treated separately (no overlap); 
\item 
linear systems can be solved independently across the quadrature nodes of the complex
contour;
\item each complex linear system with multiple right-hand-sides can be solved in parallel.
\end{enumerate}
Consequently, within a parallel environment,
{\em the algorithm complexity depends on solving a single linear
system.} 
\end{itemize}

\newpage


\subsection{The FEAST Solver Package version \version}

The FEAST numerical library package ({\bf www.feast-solver.org}) has first been developed
 and released (under free BSD license) in Sep. 2009 (v1.0), follows by upgrades in Mar. 2012 (v2.0),
and Feb. 2013 (v2.1) [version adopted by Intel-MKL]. 
The current version of the FEAST package (v3.0) released in Jun. 2015, focuses on solving the 
Hermitian and Non-Hermitian eigenvalue problems (real symmetric, real non-symmetric, complex Hermitian, complex non-Hermitian, 
complex symmetric) on both shared-memory architecture (i.e. FEAST-SMP version) and distributed architecture (i.e. FEAST-MPI version 
including the three levels of parallelism MPI-MPI-OpenMP). 

\begin{figure}[htbp]
\begin{minipage}{0.53\linewidth}
\centering
\includegraphics[width=0.9\linewidth]{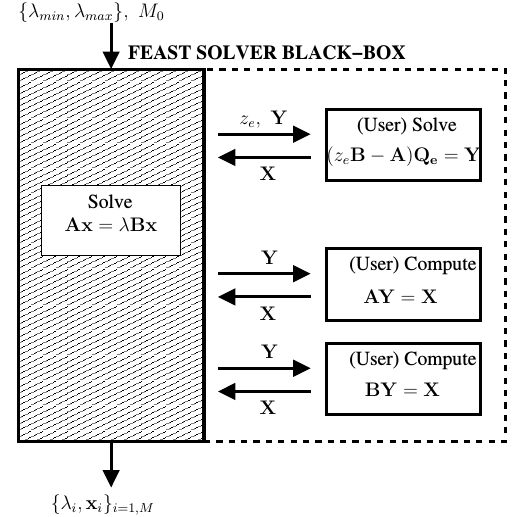}
\end{minipage}
\hfill
\begin{minipage}{0.46\linewidth}
\caption{\label{fig_bb} \small \em ``Black-box'' interface for the Hermitian problem. 
FEAST requires a search interval and a search subspace size $M_0$. 
It includes features such as reverse communication interfaces (RCI)
that are matrix format independent, and linear
system solver independent, as well as ready to
use predefined interfaces for dense, banded and
sparse systems. For the predefined interfaces case
the ``black-box'' region extends then to the right
dashed box, and only the system matrices are 
required as inputs from the users. The RCI 
interfaces represent the kernel of FEAST which can be
customized by the users to allow maximum 
flexibility for their specific applications. Users have
then the possibility to integrate their own linear
system solvers (direct or iterative – with or without preconditioner) and handle their own matrix-vector multiplication procedure.}
\end{minipage}
\end{figure}

FEAST is a comprehensive numerical library offering both simplicity and flexibility, and packaged around a
``black-box'' interface as depicted in Figure \ref{fig_bb}.

The current main features of the FEAST package include: 
\begin{itemize}
\itemsep 1pt
\parskip 1pt 
\item Standard or generalized Hermitian and non-Hermitian eigenvalue problems (left/right eigenvectors and bi-orthonormal basis); 
\item Two libraries: {\bf SMP version} (one node), and  {\bf MPI version} (multi-nodes);
\item Real/Complex arithmetic and Single/Double precisions; 
\item A set of flexible and useful practical options (quadrature rules, contour shapes,  stopping criteria, initial guess, etc.) 
\item Fast stochastic estimates for search subspace size $M_0$.
\item Source code and pre-compiled libraries provided for common architectures (e.g. x64)-  FEAST is written
in Fortran 90, but the FEAST libraries do not contain Fortran
runtime dependencies to maximize portability (e.g. compatibility with any Fortran or C compilers).
\item Reverse communication interfaces (RCI): {\bf Maximum flexibility for application specific}. {\em Those are matrix format 
independent, inner system solver independent, so users must provide their own linear system solvers (direct or iterative) and mat-vec utility routines.}  
\item Predefined driver interfaces for dense, banded, and sparse (CSR) formats: {\bf Less flexibility  but easy to use ("plug and play"):}
\begin{itemize}
\item FEAST\_DENSE interfaces require LAPACK. 
\item FEAST\_BANDED interfaces use the SPIKE-SMP linear system solver  (included)
\item FEAST\_SPARSE interfaces requires the Intel MKL-PARDISO solver.
\end{itemize} 
\item All the FEAST interfaces require (any optimized) LAPACK and BLAS packages.
\item Multiple utility routines are also included (e.g. user-defined custom contour in complex plane, 
variation of the spectral projector rational function, extract nodes/weights from predefined quadrature rules, etc.)
  \item Examples and documentation included,
\item  Utility sparse drivers included (i.e. users can also provide their matrix systems in coordinate/matrix-market format 
for testing, timing, etc.).
\end{itemize}

\vspace{0.5cm}

{\bf Remark:}
Although it could be possible to develop a large collection of FEAST drivers that can be linked with all the main linear system solver packages, 
 we are rather focusing our efforts on the development of highly efficient, functional and flexible FEAST RCI interfaces
which are placed on top of the computational hierarchy. 
Within the FEAST RCI interfaces,  maximum flexibility is indeed available to the users for choosing their preferred and/or application 
specific direct linear system method or iterative method with (customized or generic) preconditioner.


\subsubsection{Using the FEAST-SMP version}
For a given search interval, parallelism (via shared memory programming) is not explicitly implemented 
in FEAST  i.e. the inner linear systems are solved one after another within one node (avoid the fight for resources). 
Therefore,  parallelism can only be achieved if the inner system solver and the mat-vec routines are threaded. 
Using the FEAST predefined drivers, in particular, parallelism is
 implicitly obtained within the shared memory version of BLAS, LAPACK, SPIKE-SMP or MKL-PARDISO.  
If FEAST is linked with the INTEL-MKL library, the shell variable
{\tt MKL\_NUM\_THREADS} can be used for setting automatically the number of threads (cores) for both BLAS, LAPACK and MKL-PARDISO.
In the general case, the user is responsible for activated the threaded capabilities of their BLAS, LAPACK libraries and 
their own linear systems solvers - most likely using the shell variable  {\tt OMP\_NUM\_THREADS}. The latter must be defined with
 the FEAST-BANDED interfaces since they are already making use of our own SPIKE-SMP solver.  

{\bf Remark:} If memory resource is not an issue (in particular for small to moderate size systems), the flag {\tt fpm(10)}
should be changed to value 1. In this case, all the factorizations performed by the FEAST predefined drivers (dense, banded, sparse)
will be saved into memory (i.e. they are not recomputed along the FEAST iterations) and performances will improved.
With this option the
FEAST-DENSE interface, in particular, should become more competitive in comparison with the LAPACK eigenvalue routines for computing 
selected eigenpairs. 


\subsubsection{Using the FEAST-MPI version}
In the current FEAST version, only the second-level of parallelism is explicitly addressed by the code. 
This is accomplished in a trivial fashion by sending off the different linear systems 
(which can be solved independently for the points along the complex contour) along the compute nodes.
From the user perspective, interfaces and arguments list stay completely unchanged and it is the {\tt -lpfeast, etc.} library 
(instead of {\tt -lfeast, etc.})  that  needs to be linked within an MPI environment.
  Although, FEAST can run on any numbers of MPI processors, there will be  a
peak of performance if the number of MPI processes is equal to the number of contour points i.e. 
either {\tt fpm(2)} or {fpm(8)} depending on the nature the eigenvalue problem and
the FEAST drivers. 
Indeed, the MPI implementation in v3.0 does not yet provide an option for the third level of parallelism (system solver level) 
and FEAST still needs to call a shared-memory solver.
However, it is important to note that a MPI-level management has been added to allow  
 easy parallelism of the search interval using a coarser level of MPI (i.e. first level parallelism). 
For this case, a new flag {\tt fpm(9)} has been added as the only new input required by FEAST. This flag can be set equal to 
 a new local communicator variable ``MPI\_COMM\_WORLD" which contains the selected user's MPI processes for a given search interval. 
If only one search interval is used, this new flag is set by default to the global ``MPI\_COMM\_WORLD'' value in the FEAST initialization step.



\newpage 

\subsection{Installation and Setup: A Step by Step Procedure}

In this section, we address the following question: How should you install and link the FEAST library? 

\subsubsection{Installation- Compilation}

Please follow the following steps (for Linux/Unix systems):                 
\begin{enumerate}
\item Download the latest FEAST version in {\bf http://www.ecs.umass.edu/$\sim$polizzi/feast}, for example, let us call 
this package {\tt feast\_3.0.tgz}.
\item Put the file in your preferred directory such as {\tt \$HOME} directory or (for example)  {\tt /opt/} directory
 if you have ROOT privilege. 
\item Execute: {\tt \bf tar -xzvf feast\_3.0.tgz};  Figure~\ref{feast_tree} represents
the  main {\tt FEAST} tree directory being created.

\begin{figure}[htbp]
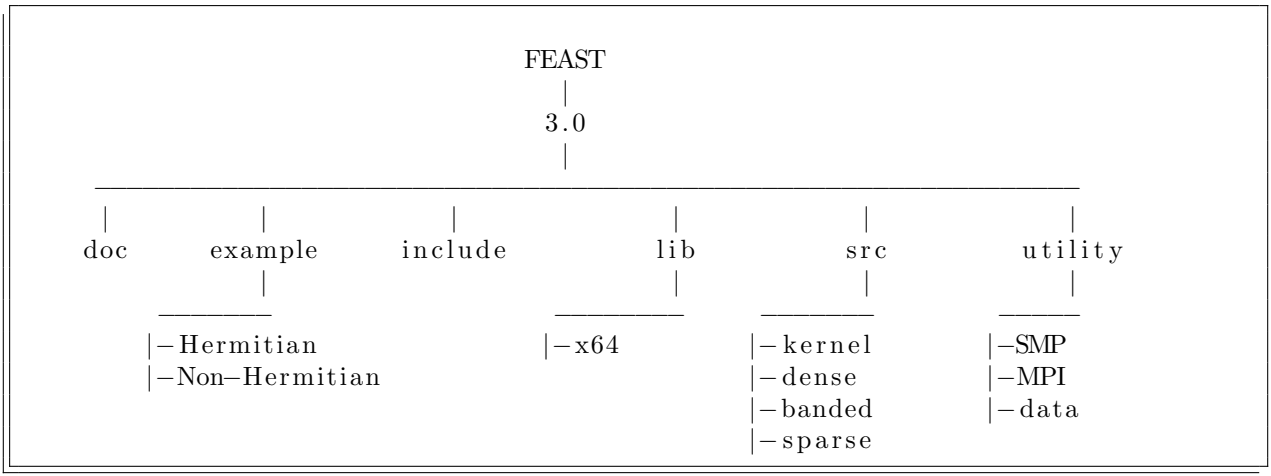

\begin{center}
\begin{lstlisting}[frame=trBL]

                                FEAST
                                  |
                                 3.0 
                                  |
     --------------------------------------------------------------
     |         |           |             |           |            |
    doc     example     include         lib         src        utility
               |                         |           |            |
         -------                  --------     -------        -----                 
        |-Hermitian              |-x64        |-kernel       |-SMP        
        |-Non-Hermitian                       |-dense        |-MPI
                                              |-banded       |-data
                                              |-sparse

\end{lstlisting}
\caption{\label{feast_tree} Main FEAST tree directory.}
\end{center}
\end{figure}

\item let us denote \feastdir  the package's main directory after
installation. For example, it could be
\begin{quote}
{\tt $\sim$/home/FEAST/3.0}~~~ or ~~~{\tt /opt/FEAST/3.0}.
\end{quote}
{\bf It is not mandatory but recommended to define the Shell variable {\tt \$FEASTROOT},} e.g.  
\begin{quote}
{\tt \bf export FEASTROOT=\feastdir} \\ or~~~{\tt \bf set FEASTROOT=\feastdir} 
\end{quote}
respectively  for the BASH or CSH shells.
One of this command can be placed in the appropriate shell startup file in {\tt \$HOME} (i.e {\tt .bashrc}
 or {\tt .cshrc}).
\item The FEAST pre-compiled libraries can be found at
\begin{quote}
{\tt \$FEASTROOT/lib/\arch}
\end{quote}
where \arch denotes the computer architecture. Currently, the following architectures are provided:
\begin{itemize}
\item {\tt x64} for common 64 bits architectures 
(e.g. Intel em64t: Nehalem, Xeon, Pentium, Centrino etc.;  amd64),
\end{itemize}
The pre-compiled libraries are free from Fortran90 runtime dependencies (i.e. they can be called from any {\tt Fortran} or {\tt C} codes
without compatibility issues). For the FEAST-MPI, the precompiled library include two versions  MPICH2 and OpenMPI.
If your current architecture is listed above, you can proceed directly to step {\bf 7}, if not,
 you will need to compile the FEAST library in the next step. You would also need to compile FEAST-MPI if you are using a different 
MPI implementation that the one proposed here. 
\item {\it Compilation of the FEAST library source code}:

\begin{itemize}
\item Go to the directory {\tt \$FEASTROOT/src}
\item {\bf Edit the {\tt make.inc} file, and follow the directions.} Depending on your options, you would need to change appropriately the name/path 
of the {\tt Fortran90} or/and {\tt C Compilers} and optionally {\tt MPI}.

Two main options are possible:
\begin{enumerate}
\item[1-] FEAST is written in {\tt Fortran90} so direct compilation is possible using any {\tt Fortran90} compilers 
(tested with {\tt ifort} and {\tt gfortran}). 
If this option is selected, users must then be aware of runtime
 dependency problems. For example, if the FEAST library is compiled using {\tt ifort} but the user code is compiled using {\tt gfortran} or 
{\tt C} then the flag {\tt -lifcoremt} should be added to this latter; In contrast, if the FEAST library is compiled using {\tt gfortran} but the user 
code is compiled using {\tt ifort} or {\tt C}, the flag {\tt -lgfortran} should be used instead. 
\item[2-] Compilation free from Fortran90 runtime dependencies (i.e. some low-level Fortran intrinsic functions are replaced by C ones). 
This is the best option since once compiled, the library could be called from any {\tt Fortran} or {\tt C} codes
removing compatibility issues. This compilation can be performed using any {\tt C} compilers ({\tt gcc} for example), 
 but it currently does require the use of the Intel Fortran Compiler. 
\end{enumerate}

The same source codes are used for compiling FEAST-SMP and/or FEAST-MPI. For this latter, 
the MPI instructions are then activated by compiler directives (a flag {\tt <-DMPI>} is added).
The user is also free to choose any MPI implementations (tested with  Intel-MPI, MPICH2 and OpenMPI).  

\item For creating the FEAST-SMP: Execute:\\ {\tt \bf make ARCH=\arch LIB=feast all}\newline
 where \arch is your selected name for your architecture; your FEAST library including: \\
{\tt libfeast\_sparse.a \\libfeast\_banded.a \\libfeast\_dense.a \\libfeast.a} \\
will then be created in {\tt \$FEASTROOT/lib/\arch}.
\item Alternatively, for creating the FEAST-MPI library:  Execute: \\ {\tt \bf make ARCH=\arch LIB=pfeast all} to obtain: \\
{\tt libpfeast\_sparse.a \\libpfeast\_banded.a \\libpfeast\_dense.a \\libpfeast.a} \\
You may want to rename these libraries with a particular extension name associated with your MPI compilation.

\end{itemize}
\item Congratulations, FEAST is now installed successfully on your computer !!

\newpage

\subsubsection{Linking FEAST}

In order to use the FEAST library for your {\tt F77}, {\tt F90}, {\tt C} or {\tt MPI} application, you will then need to add the following instructions
in your {\tt Makefile}:
\begin{itemize}
\item {\it for the LIBRARY PATH:}~~ {\tt -L/\$FEASTROOT/lib/\arch } 
\item {\it for the LIBRARY LINKS using FEAST-SMP:} {\bf (examples)}\\
                   {\tt -lfeast}    (FEAST kernel alone - Reverse Communication Interfaces)\\
                   {\tt -lfeast\_dense -lfeast}  (FEAST dense interfaces)\\
                   {\tt -lfeast\_banded -lfeast} (FEAST banded interfaces)\\
                   {\tt -lfeast\_sparse -lfeast} (FEAST sparse interfaces)\\
                   {\tt -lfeast\_sparse -lfeast\_banded -lfeast} (FEAST sparse and banded interfaces)
\item {\it for the LIBRARY LINKS using FEAST-MPI:}{\bf (examples)}\\
                   {\tt -lpfeast<ext>}    (FEAST kernel alone - Reverse Communication Interfaces)\\
                   {\tt -lpfeast\_dense<ext> -lpfeast<ext>}  (FEAST dense interfaces)\\
                   {\tt -lpfeast\_banded<ext> -lpfeast<ext>} (FEAST banded interfaces)\\
                   {\tt -lpfeast\_sparse<ext> -lpfeast<ext>} (FEAST sparse interfaces)\\
                   {\tt -lpfeast\_sparse<ext> -lpfeast\_banded<ext> -lpfeast<ext>} (FEAST sparse and banded interfaces)\\
where, in the precompiled library,  {\tt <ext>} is the extension name associated with {\tt \_impi}, {\tt \_mpich2} or {\tt \_openmpi} respectively
for Intel MPI, MPICH2 and OpenMPI. \\

In order to illustrate how should one use the above FEAST library links including dependencies,  
let us call (for example) 
 {\tt -llapack}, {\tt -lblas}
 respectively your link for the your optimized LAPACK and BLAS packages. The complete library links with dependencies
are then given for FEAST-SMP or FEAST-MPI by {\bf (examples)}:\\ 
  {\tt -l<p>feast -l<yourownsystemsolver> -llapack -lblas} \\
             {\tt  -l<p>feast\_dense  -lfeast -llapack -lblas} \\
             {\tt  -l<p>feast\_banded -lfeast -llapack -lblas} \\

{\bf Remarks} \\
1- {\tt -l<yourownsystemsolver>} represents the link to your own system solver in Figure \ref{fig_bb}. \\
2- If {\tt -lfeast\_sparse} or  {\tt  -lpfeast\_sparse<ext>} are used, they must be linked with Intel MKL (which contains 
both MKL\_PARDISO, LAPACK and BLAS) \\[10pt]

\item  {\it for the INCLUDE PATH:}~~  {\tt  -I/\$(FEASTROOT)/include} \\
It is mandatory only for {\tt C} codes. Additionally, instructions need to be added in the header {\tt C} file {\bf (all that apply)}:  \\ 
{\tt \#include "feast.h"} \\
{\tt \#include "feast\_sparse.h"} \\
{\tt \#include "feast\_banded.h"} \\
{\tt \#include "feast\_dense.h"}

\end{itemize}

\end{enumerate}

\newpage

\subsection{A simple ``Hello World'' Example ({\tt F90, C, MPI-F90, MPI-C})}\label{sec_hello}

This example solves a 2-by-2 dense standard eigenvalue system $\bf Ax=\lambda x$ where

\begin{equation}
{\bf A}=
\begin{pmatrix}
 2 & -1 \\
-1 & 2  
 \end{pmatrix}
\end{equation}
and the two eigenvalue solutions are known to be $\lambda_1=1$, $\lambda_2=3$ which can be associated
respectively with the orthonormal eigenvectors $(\sqrt 2/2,\sqrt 2/2)$ and $(\sqrt 2/2,-\sqrt 2/2)$.

Let us suppose that one can specify a search interval,
a single call to the {\tt DFEAST\_SYEV} subroutine solves then this dense standard eigenvalue system 
in double precision. Also, the FEAST parameters can be set to their default values by 
a call to the {\tt FEASTINIT} subroutine.

\subsubsection*{F90}

\begin{figure}[htbp]
\begin{footnotesize}
\begin{center} \includegraphics[width=0.78\linewidth,angle=0]{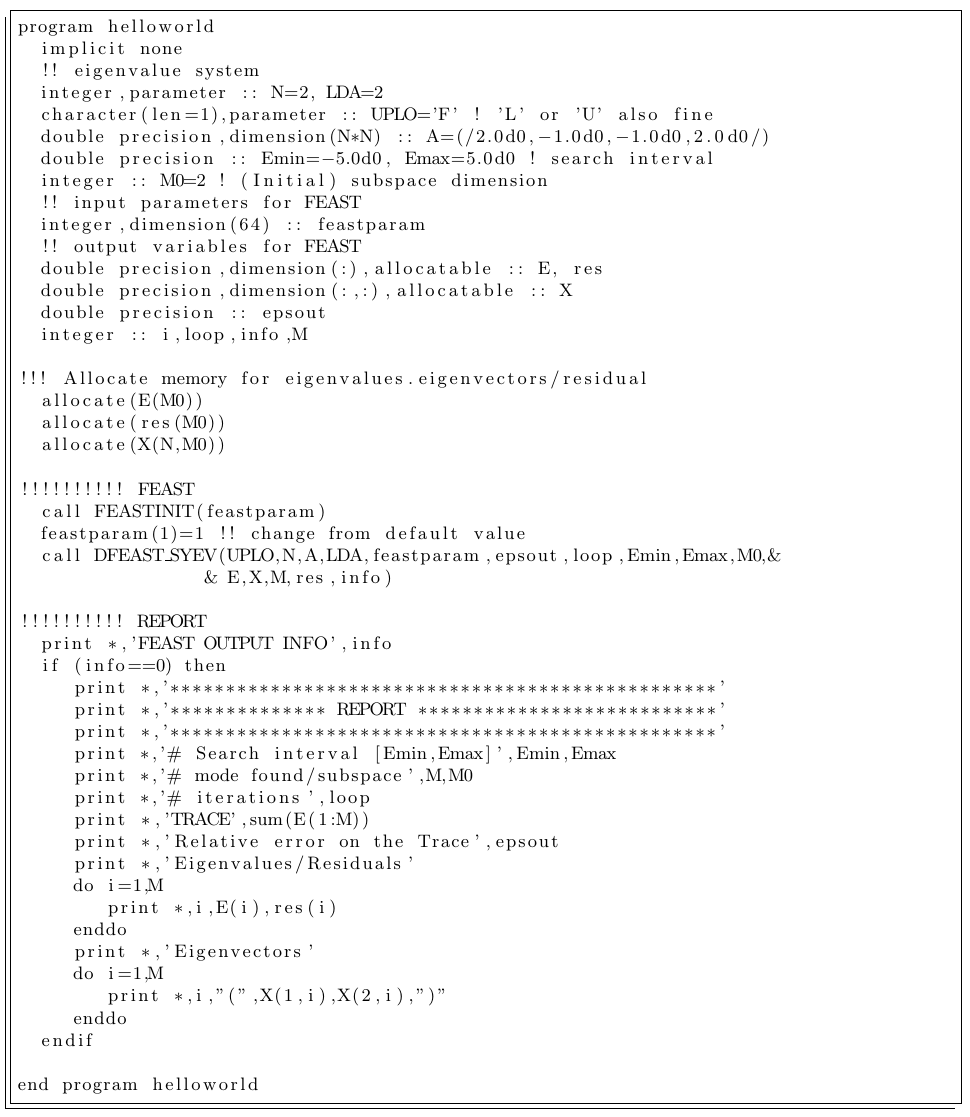}\end{center}
\caption{A very simple F90 ``helloworld'' example.
 This code can be found in {\tt \feastdir/example/Hermitian/Fortran/1\_dense}.}
\label{figure:hello_world}
\end{footnotesize}
\end{figure}

The Fortran90 source code of {\tt helloworld.f90} is listed in Figure~\ref{figure:hello_world}.

To create the executable, compile and link the source program with the
FEAST \version library, one can use:\\

\noindent
 {\tt >ifort helloworld.f90 -o helloworld -L<{\it {\tt FEAST} directory}>/lib/<{\it arch}>   -lfeast\_dense -lfeast -mkl} \\
where we assume that: (i) the FORTRAN compiler is {\tt ifort} the Intel's one,
(ii) the FEAST package has been installed in a directory called {\tt <{\it {\tt FEAST} directory}>},
(iii) the user architecture is {\tt <{\it arch}>} (x64, ia64, etc.), (iv) MKL is used to link the LAPACK and BLAS libraries.

A run of the resulting executable looks like $${\tt>./helloworld}$$
and the output of the run appears in Figure~\ref{figure:hello_worldout}.

\begin{figure}[htbp]
\begin{footnotesize}
\input{feast_helloworld_out}
\caption{Output results for the simple F90 ``helloworld'' example.}
\label{figure:hello_worldout}
\end{footnotesize}
\end{figure}

\newpage

\subsubsection*{C}

\begin{figure}[htbp]
\begin{footnotesize}
\centering \includegraphics[width=0.72\linewidth,angle=0]{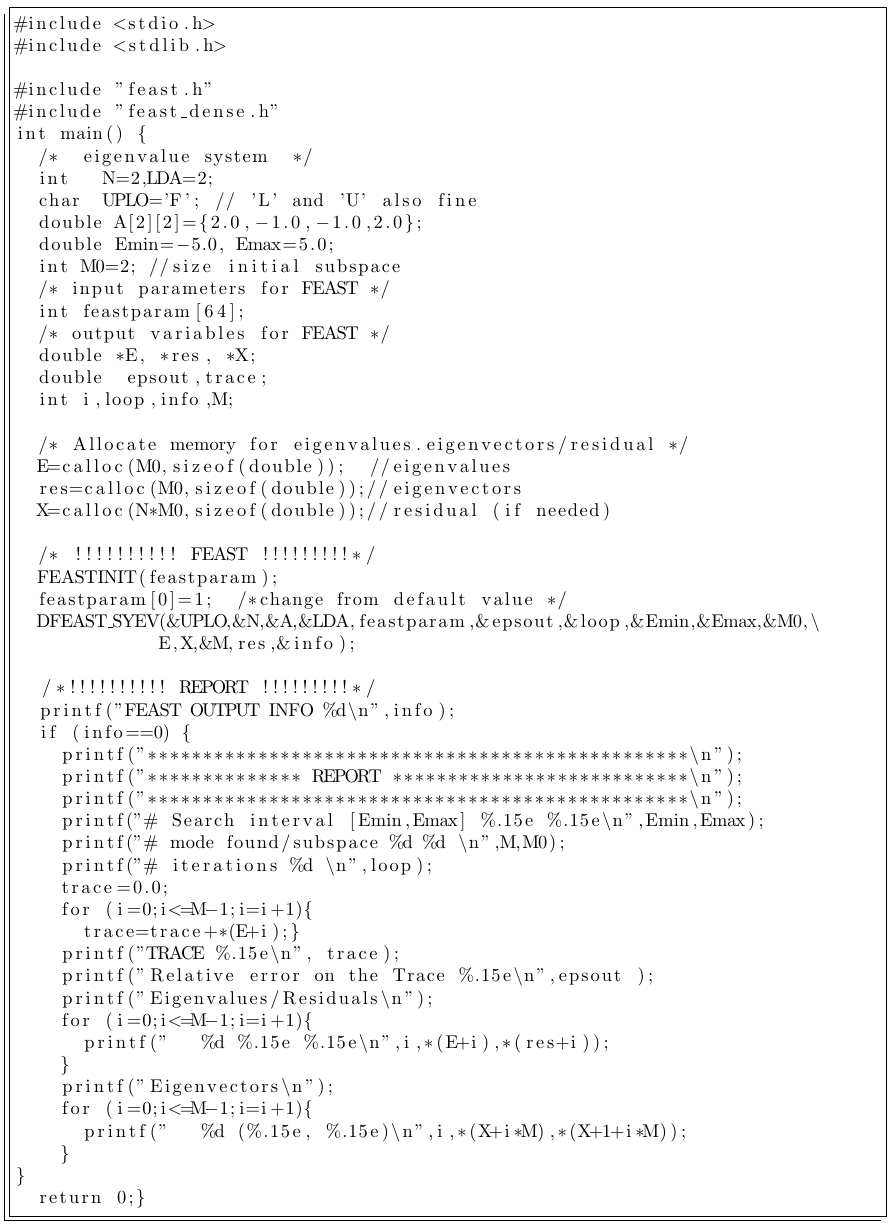}
\caption{A very simple C ``helloworld'' example. This code can be found in {\tt \feastdir/example/Hermitian/C/1\_dense}.}
\label{figure:hello_worldC}
\end{footnotesize}
\end{figure}

Similarly to the F90 example, the corresponding C source code of the {\tt helloworld} example ({\tt helloworld.c}) is listed
in Figure~\ref{figure:hello_worldC}. The executable can now be created using the {\it gcc} compiler (for example),
along with the {\tt -lm} library: 

  {\tt gcc helloworld.c -o helloworld \BS \\   
      \Indent -I<{\it {\tt FEAST} directory}>/include -L<{\it {\tt FEAST} directory}>/lib/<{\it arch}>   -lfeast\_dense -lfeast \BS \\
      \Indent -Wl,--start-group  -lmkl\_intel\_lp64 -lmkl\_intel\_thread -lmkl\_core  -Wl,--end-group \BS \\ 
      \Indent       -liomp5 -lpthread -lm} \\
where we assume that FEAST has been compiled without runtime dependencies. In contrast, if the FEAST library was compiled using {\tt ifort} alone 
  then the flag {\tt -lifcoremt} should be added above; In turn, if the FEAST library was compiled using {\tt gfortran} alone, it is
 the flag {\tt -lgfortran} that should be added instead.

\newpage

\subsubsection*{MPI-F90}

Similarly to the F90 example, the corresponding MPI-F90 source code of the {\tt helloworld} example ({\tt phelloworld.f90}) is listed
in Figure~\ref{figure:hello_worldMPI}. The executable can now be created using {\tt mpif90} (for example):

  {\tt mpif90 -f90=ifort phelloworld.f90 -o phelloworld \BS \\   
      \Indent -L<{\it {\tt FEAST} directory}>/lib/<{\it arch}>   -lpfeast\_dense -lpfeast -mkl} \\ 
where we assume that: (i) the Intel Fortran compiler is used, (ii) the FEAST-MPI library has been compiled using 
the same MPI implementation.

A run of the resulting executable looks like $${\tt>mpirun -ppn~1 -n <x> ./phelloworld}$$

where ${\tt <x>}$ represents the number of nodes.

\subsubsection*{MPI-C}

\begin{figure}[htbp]
\begin{footnotesize}
\includegraphics[width=0.5\linewidth,angle=0]{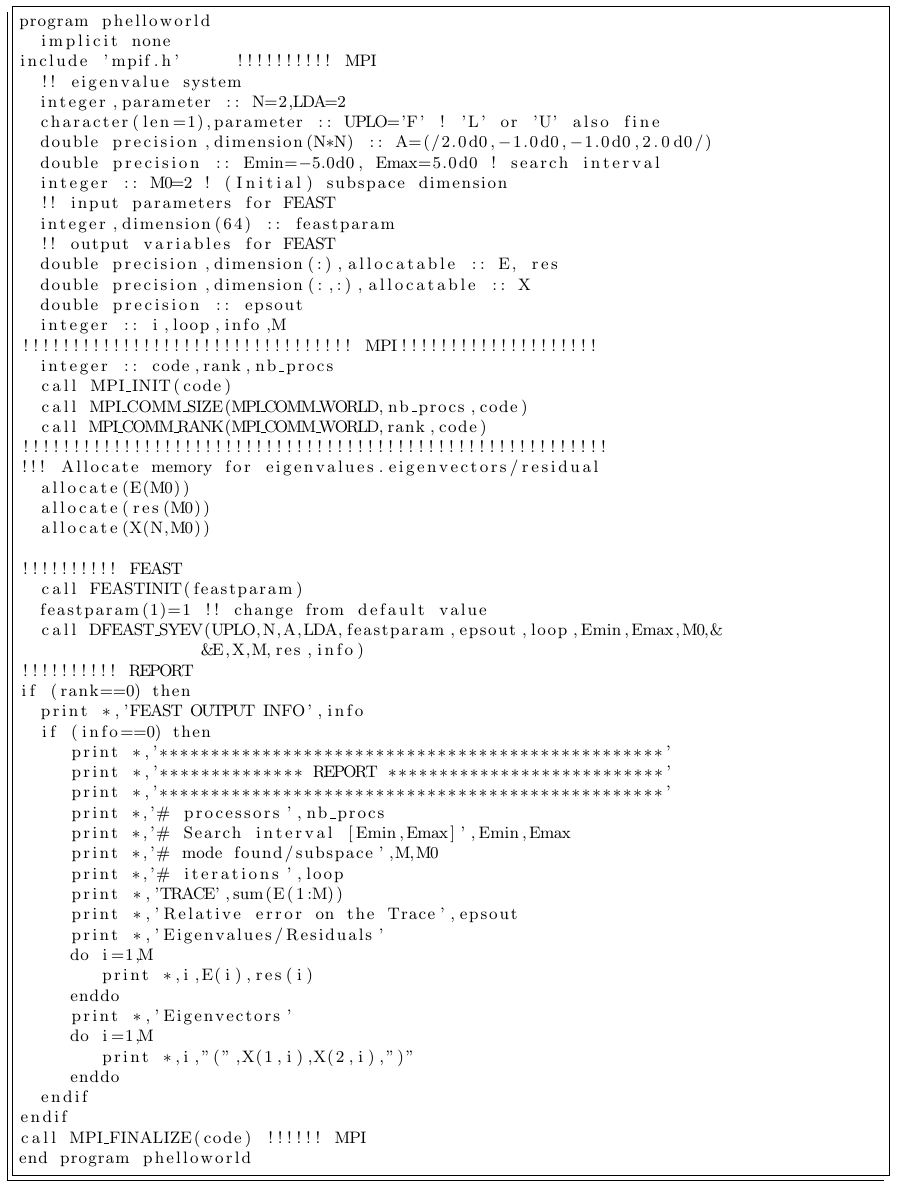}
\hfill
\includegraphics[width=0.5\linewidth,angle=0]{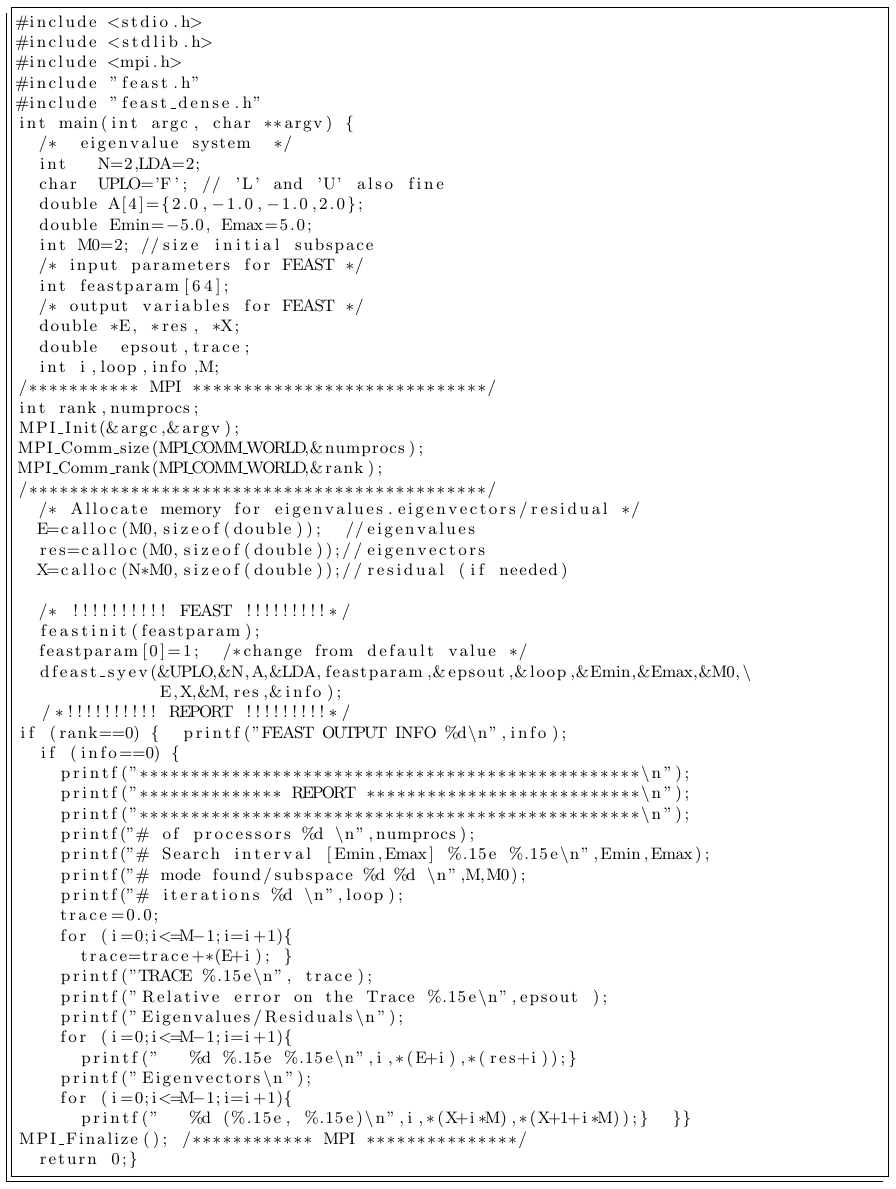}
\caption{A very simple MPI-F90 and MPI-C ``helloworld'' example.
These codes can be found in {\tt \feastdir/example/Hermitian/<Fortran-MPI,C-MPI>/1\_dense}} 
\label{figure:hello_worldMPI}
\end{footnotesize}
\end{figure}

Similarly to the MPI-F90 example, the corresponding MPI-C source code of the {\tt helloworld} example ({\tt phelloworld.c}) is listed
in Figure~\ref{figure:hello_worldMPI}. The executable can now be created using {\tt mpicc} (for example):

  {\tt mpicc -cc=gcc helloworld.f90 -o helloworld \BS \\   
      \Indent -L<{\it {\tt FEAST} directory}>/lib/<{\it arch}>   -lpfeast\_dense -lpfeast \BS \\
      \Indent -Wl,--start-group  -lmkl\_intel\_lp64 -lmkl\_intel\_thread -lmkl\_core  -Wl,--end-group \BS \\ 
      \Indent       -liomp5 -lpthread -lm} \\
where we assume that: (i) the gnu {\tt C} compiler is used, (ii) the FEAST-MPI library has been compiled using the same MPI implementation
 , (iii) FEAST has been compiled without runtime dependencies (otherwise see comments in {\tt C} example section).


%% file: feast_helloworld_out.tex
\begin{footnotesize}
\begin{lstlisting}[frame=trBL]
***********************************************
*********** FEAST- BEGIN **********************
***********************************************
Routine DFEAST_S{}{}
List of input parameters fpm(1:64)-- if different from default
   fpm(1)=1
Search interval [-5.000000000000000e+00; 5.000000000000000e+00]
Size system    	2
Size subspace  	2
#Linear systems	8
-----------------------------------------------------------------------------------
#Loop | #Eig  |       Trace           |     Error-Trace       |     Max-Residual
-----------------------------------------------------------------------------------
0	2	4.000000000000015e+00	1.000000000000000e+00	2.512147933894036e-15
==>FEAST has successfully converged (to desired tolerance)
***********************************************
*********** FEAST- END*************************
***********************************************

 FEAST OUTPUT INFO           0
 *************************************************
 ************** REPORT ***************************
 *************************************************
 # Search interval [Emin,Emax]  -5.00000000000000        5.00000000000000     
 # mode found/subspace           2           2
 # iterations           0
 TRACE   4.00000000000002     
 Relative error on the Trace   1.00000000000000     
 Eigenvalues/Residuals
           1   1.00000000000000       7.536443801682115E-016
           2   3.00000000000001       2.512147933894036E-015
 Eigenvectors
           1 (  0.707106781186547       0.707106781186549      )
           2 (  0.707106781186552      -0.707106781186545      )
\end{lstlisting}
\end{footnotesize}

%% file: feast_interfaces.tex
\section{FEAST Interfaces}


\subsection{Basics}

\subsubsection{Definition}

There are two different type of interfaces available in the FEAST library:\\

\noindent \fbox{
\begin{minipage}[t][][b]{0.49\linewidth}
{\bf Reverse communication interfaces (RCI):}
                    $${\tt  {\bf \large T}feast\_{\bf \large Y}rci} $$
with their expert version
$${\tt  {\bf \large T}feast\_{\bf \large Y}rcix}$$
These interfaces constitute the kernel of FEAST. They are 
matrix free format (the interfaces are independent of the matrix data formats), 
users can then define their own explicit or implicit data format. Mat-vec routines and
direct/iterative linear system solvers must also be provided by the users. \\
\end{minipage}}
 \hfill
\fbox{
\begin{minipage}[t][][b]{0.49\linewidth}
{\bf Format predefined interfaces:}
                    $${\tt  {\bf T}feast\_{\bf \large YF}ev} \quad \mbox{ and } \quad {\tt  {\bf \large T}feast\_{\bf \large YF}gv},$$
with their expert versions
               $${\tt  {\bf T}feast\_{\bf \large YF}evx} \quad \mbox{ and } \quad {\tt  {\bf \large T}feast\_{\bf \large YF}gvx},$$
These interfaces for standard ``{\tt ev}'' and generalized ``{\tt gv}'' problems can be considered as predefined optimized drivers for {\tt {\bf \large T}feast\_{\bf \large Y}rci} and  {\tt {\bf \large T}feast\_{\bf \large Y}rcix} that
 act on commonly used matrix data storage (dense, banded and sparse-CSR), using
predefined mat-vec routines and preselected  inner linear system solvers.
\end{minipage}}

\begin{itemize}
\item {\tt \bf T} is the data type of matrix {\bf A} (and matrix {\bf B} if any) i.e.
\begin{small}
\begin{tabular}{|l|l|} \hline
  Value of {\tt {\bf T}} & Type of matrices \\ \hline\hline
 {\tt s} & single precision \\ \hline 
 {\tt d} & double precision \\ \hline
 {\tt c} & complex single precision \\ \hline
 {\tt z} &  complex double precision \\ \hline
\end{tabular}
\end{small}

\item {\tt \bf \large YF} is the problem type (Symmetric, Hermitian, or General) and storage format (Dense, Banded, Sparse).
Together they define the problem.
\vspace{-.25cm}
\begin{center}
\begin{footnotesize}
\begin{tabular}{|c|c|c|c|c|}             \hline
 {\tt \bf Y} &  {\tt \bf F} & Problem Type & Matrix Format & Linear Solver \\
\hline
\hline
 s &  y & Symmetric & Dense & LAPACK \\ \hline
 h &  e & Hermitian & Dense & LAPACK \\ \hline
 g &  e & General   & Dense & LAPACK \\ \hline
 s &  b & Symmetric & Banded & SPIKE \\ \hline
 h &  b & Hermitian & Banded & SPIKE  \\ \hline
 g &  b & General   & Banded & SPIKE  \\ \hline
 s &  csr & Symmetric & Sparse & MKL-PARDISO \\ \hline
 h &  csr & Hermitian & Sparse & MKL-PARDISO \\ \hline
 g &  csr & General & Sparse & MKL-PARDISO \\ \hline
\end{tabular}
\end{footnotesize}
\end{center} 
\end{itemize}

The different FEAST interface names and combinations are summarized in Table \ref{feast_int}.

\begin{table}[htbp]
\begin{center}
\begin{small}
\begin{tt}
\begin{tabular}{|c|}
    \multicolumn{1}{l}{RCI- interfaces} \\ \hline 
       \{s,d,c,z\}feast\_srci \\ 
       \{c,z\}feast\_hrci \\
       \{s,d,c,z\}feast\_grci \\ \hline 
    \multicolumn{1}{l}{DENSE- interfaces} \\ \hline 
      \{s,d,c,z\}feast\_sy\{ev,gv\} \\
      \{c,z\}feast\_he\{ev,gv\} \\
      \{s,d,c,z\}feast\_ge\{ev,gv\} \\ \hline 
\end{tabular}
\begin{tabular}{|c|c|}
    \multicolumn{1}{l}{BANDED- interfaces} \\ \hline 
     \{s,d,c,z\}feast\_sb\{ev,gv\} \\ 
     \{c,z\}feast\_hb\{ev,gv\} \\ 
     \{s,d,c,z\}feast\_gb\{ev,gv\} \\ \hline 
    \multicolumn{1}{l}{SPARSE- interfaces} \\ \hline 
      \{s,d,c,z\}feast\_scsr\{ev,gv\} \\ 
      \{c,z\}feast\_hcsr\{ev,gv\} \\ 
      \{s,d,c,z\}feast\_gcsr\{ev,gv\} \\ \hline 
\end{tabular}
\end{tt}
\caption{\label{feast_int} List of all FEAST interfaces available in FEAST v3.0.  Expert routines 
include {\tt x} at the end.}
\end{small}
\end{center}
\end{table}


\subsubsection{Common Declarations}
The arguments list for the FEAST interfaces are commonly  defined as follows: 
\begin{quote}
\begin{tt}
{\bf T}feast\_{\bf Y}{\it \{interface\}}~ ({\bf \{List\}},fpm,epsout,loop,{\bf \{List-I\}},M0,E,X,M,res,info)\\
{\bf T}feast\_{\bf Y}{\it \{interface\}}x ({\bf \{List\}},fpm,epsout,loop,{\bf \{List-I\}},M0,E,X,M,res,info,Zne,Wne)
\end{tt}
\end{quote}

where {\bf\{List\}} and {\bf\{List-I\}} denote a series of arguments that are specific to each interfaces and will be 
presented in the next sections. The rest of the arguments are common to both RCI and predefined FEAST interfaces
and their definition is given in Table \ref{tab_common}.

\begin{table}[htbp]
\begin{center}
\begin{footnotesize}
\begin{tabular}{|l|l|c|l|}
\hline
& & & \\ 
 & Type (Fortran) & Input/Output &Description \\ 
 &  & & \\ \hline  \hline
{\tt fpm}  & 
integer(64)  & in & FEAST input parameters (see Table \ref{tab_fpm})\\ 
& & & The size should be at least 64 \\ \hline
{\tt epsout}  & Type(S) \hfill {\it if {\bf T}={\tt s,c}}  & out & Relative error on the trace \\
              & {\it or}                                       &        &  {\footnotesize $|{\tt trace}_k-{\tt trace}_{k-1}|/\max({\tt |Emin|,|Emax|})$}, or  \\
              & Type(D) \hfill {\it if {\bf T}={\tt d,z}} &  &  {\footnotesize $|{\tt trace}_k-{\tt trace}_{k-1}|/({\tt |Emid|+r})$}\\ \hline
{\tt loop} &  integer           &  out            &    \# of FEAST subspace iterations                      \\ \hline
{\tt M0} &    integer         &    in/out          &  Search subspace dimension               \\ 
         &                    &                          & On entry: initial guess ({\tt M0>M})        \\
         &                    &                          & On exit:  new suitable {\tt M0} if guess too large    \\ \hline
{\tt E} &    Type(S)({\tt M0}) \hfill {\it if {\bf TY}={\tt ss,ch}}           &     out         & Eigenvalues     \\
              & Type(D)({\tt M0}) \hfill {\it if {\bf TY}={\tt ds,zh}} & &  the first {\tt M} values are in the search interval  \\ 
              & Type(C)({\tt M0}) \hfill {\it if {\bf TY}={\tt sg,cg,cs}} & & the others {\tt M0-M} values are outside \\ 
              & Type(Z)({\tt M0}) \hfill {\it if {\bf TY}={\tt dg,zg,zs}} & &                                         \\ \hline
{\tt X} &    Type(S)({\tt N,M0}) \hfill {\it if {\bf TY}={\tt ss}}        &     in/out       
  & Eigenvectors ({\tt N}: size of the system)                 \\
            &  Type(D)({\tt N,M0}) \hfill {\it if {\bf TY}={\tt ds}}    & & On entry: guess subspace if {\tt fpm(5)=1}  \\ 
& Type(C)({\tt N,M0}) \hfill {\it if {\bf TY}={\tt ch,cs}}  & & On exit: (right) eigenvectors solutions {\tt X(1:N,1:M)}\\
& Type(Z)({\tt N,M0}) \hfill {\it if {\bf TY}={\tt zh,zs}} & & (same ordering as in {\tt E})\\ 
& Type(C)({\tt N,2*M0}) \hfill {\it if {\bf TY}={\tt sg,cg}} & &Remark: * left vectors (if calculated) in {\tt X(1:N,M0+1:M0+M)}  \\
& Type(Z)({\tt N,2*M0}) \hfill {\it if {\bf TY}={\tt dg,zg}} & &~~~~~~~~~~~~~* if {\tt fpm(14)=1}, first $\bf Q$ subspace on exit  \\ \hline
{\tt M} &     integer        &     out         & \# Eigenvalues found in search interval    \\ 
 &             &             & \# estimated eigenvalues if {\tt fpm(14)=2} \\  \hline
{\tt res} &    Type(S)({\tt M0}) \hfill {\it if {\bf TY}={\tt ss,ch,cs}}          &   out     & Relative residual {\tt res(1:M)} (right); 
{\tt res(M0+1:M0+M)} (left)     \\
      & Type(D)({\tt M0}) \hfill {\it if {\bf TY}={\tt ds,zh,zs}}       &        &   (right) $\bf ||Ax_i - \lambda_i Bx_i ||_1 /||\alpha Bx_i ||_1$  \\
              & Type(S)({\tt 2*M0}) \hfill {\it if {\bf TY}={\tt sg,cg}} & &  (left) $\bf ||A^Hx_i - \lambda_i^* B^Hx_i ||_1 /||\alpha B^Hx_i ||_1$  
 \\ 
              &Type(D)({\tt 2*M0}) \hfill {\it if {\bf TY}={\tt dg,zg}} & & Remark: * ${\tt \alpha}=\max({\tt |Emin|,|Emax|})$ or 
${\tt \alpha}=({\tt |Emid|+r})$ \\ 
  &  & & ~~~~~~~~~~~~~* if {\tt fpm(14)=2}, {\tt res(1:M)} running average for {\tt M} \\ 
\hline
{\tt info} &  integer        &  out           &   Error handling  (if =0: successful exit)     \\ 
 & & & (see Table \ref{tab_info} for all INFO return codes) \\ \hline
{\tt Zne,Wne}  &   Type(C)({\tt fpm(2)})  \hfill {\it if {\bf TY}={\tt ss,ch}}   &  in  & Custom integration nodes and weights- Expert mode\\   
            &   Type(Z)({\tt fpm(2)})  \hfill {\it if {\bf TY}={\tt ds,zh}}      &   &   \\
            &   Type(C)({\tt fpm(8)})  \hfill {\it if {\bf TY}={\tt sg,cg,cs}}      &   &   \\   
            &   Type(Z)({\tt fpm(8)})  \hfill {\it if {\bf TY}={\tt dg,zg,zs}}     &   &   \\ \hline
\end{tabular}
\caption{\label{tab_common} List of arguments common for the RCI and predefined FEAST interfaces. \newline
{\bf Remark:} the arrays {\tt E}, {\tt X} and {\tt res} return the eigenpairs and associated residuals. 
The solutions within the intervals are contained in the first {\tt M} components of the arrays. The left vectors (if calculated) are contained in {\tt X(1:N,M0+1:M0+M)}. 
Note for expert use: the solutions that are directly outside the intervals can also be found with less accuracy 
in the other {\tt M0-M} components  (i.e. from element {\tt M+1} to {\tt M0}). In addition where spurious solutions
may be found in the processing of the FEAST algorithm, those are put at the end of the arrays {\tt E} and {\tt X} and 
are flagged with the value $-1$ in the array {\tt res}.
}
\end{footnotesize}
\end{center}
\end{table}

\newpage


\subsection{FEAST\_RCI interfaces}


These interfaces are useful if your application requires specific linear system solvers (direct or iterative) 
or/and specific matrix storage (explicit or implicit). If this is not the case,
you may want to skip this section and go directly to the section \ref{sec_feast_pre} on 
predefined interfaces.

\subsubsection{Specific declarations}

The arguments list for the FEAST\_RCI interfaces is defined as follows: 
\begin{quote}
\begin{tt}
{\bf \large T}feast\_{\bf  \large Y}rci ({\bf \{List-{rci}\}},fpm,epsout,loop,{\bf \{List-I\}},M0,E,X,M,res,info)
\end{tt}
\end{quote}
\begin{quote}
\begin{tt}
{\bf \large T}feast\_{\bf  \large Y}rcix ({\bf \{List-{rci}\}},fpm,epsout,loop,{\bf \{List-I\}},M0,E,X,M,res,info,Zne,Wne) 
\end{tt}
\end{quote}

The series  of arguments in {\bf\{List-{rci}\}} and {\bf\{List-{I}\}} are defined 
in Table \ref{tab_pre} and their description is provided 
in Table \ref{tab_rci}.

\begin{table}[htbp]
\begin{center}
\begin{small}
\begin{tabular}{|c|c||c|c|}
\hline
\tt \bf T & \tt \bf Y & {\tt\bf  List-rci} & {\tt \bf List-I}  \\
\hline
\hline
\tt d,s & \tt s & \tt  \{ijob,N,Ze,work1,zwork2,Aq,Bq\} &\tt  \{Emin,Emax\} \\ \hline
\tt z,c & \tt h & \tt  \{ijob,N,Ze,zwork1,zwork2,zAq,zBq\} &\tt  \{Emin,Emax\} \\ \hline
\tt d,s & \tt g & \tt  \{ijob,N,Ze,zwork1,zwork2,zAq,zBq\} &\tt  \{Emid,r\} \\ \hline
\tt z,c & \tt s & \tt  \{ijob,N,Ze,zwork1,zwork2,zAq,zBq\} &\tt  \{Emid,r\} \\ \hline
\tt z,c & \tt g & \tt  \{ijob,N,Ze,zwork1,zwork2,zAq,zBq\} &\tt  \{Emid,r\} \\ \hline
\hline
\end{tabular}
\end{small}
\caption{\label{tab_pre} List of arguments specific for the {\tt {\bf \large T}feast\_{\bf  \large Y}rci\{x\}} interfaces.}
\end{center}
\end{table}

\begin{table}[htbp]
\begin{center}
\begin{footnotesize}
\begin{tabular}{|l|l|c|l|}
\hline
& & & \\ 
 & Type (Fortran) & Input/ &Description \\
& & Output & \\ 
 &  & & \\ \hline  \hline
{\tt ijob}  & 
integer  & in/out & ID of the FEAST\_RCI operation \\ 
& & & On entry: ijob=-1 (initialization) \\
& & & On exit: ijob=0,10,11,20,21,30,31,40,41 \\ \hline
{\tt N}  & integer  & in & Size of the system \\ \hline
{\tt Ze} &  Type(C)  \hfill {\it if {\bf T}={\tt s,c}}   &  out    &  Coordinate along the complex contour  \\ 
      & Type(Z) \hfill {\it if {\bf T}={\tt d,z}} & & \\ \hline
{\tt work1} &   Type(S)({\tt N,M0})  \hfill {\it if {\bf T}={\tt s}}  &     in/out       
  & Workspace               \\
            & Type(D)({\tt N,M0})  \hfill {\it if {\bf T}={\tt d}}  & &  \\ \hline 
{\tt zwork1} &  
Type(C)({\tt N,M0})   \hfill {\it if {\bf TY}={\tt ch,cs}}   &  in/out    &  Workspace  \\ 
& Type(C)({\tt N,2*M0}) \hfill {\it if {\bf TY}={\tt sg,cg}} & & \\ 
      & Type(Z)({\tt N,M0}) \hfill {\it if {\bf TY}={\tt zh,zs}} & & \\ 
& Type(Z)({\tt N,2*M0}) \hfill {\it if {\bf TY}={\tt dg,zg}} & & \\ \hline
{\tt zwork2} &  Type(C){\tt N,M0})  
 \hfill {\it if {\bf T}={\tt s,c}}   &  in/out    &  Workspace  \\ 
      & Type(Z)({\tt N,M0}) \hfill {\it if {\bf T}={\tt d,z}} & & \\ \hline
{\tt Aq} {\it or} {\tt Bq} &    Type(S){\tt (M0,M0)}  \hfill {\it if {\bf T}={\tt s}}   &     in/out       
  & Workspace for the reduced eigenvalue problem         \\
            &  Type(D){\tt (M0,M0)}  \hfill {\it if {\bf T}={\tt d}}    & &  \\ \hline 
{\tt zAq} {\it or} {\tt zBq} &    Type(C){\tt (M0,M0)}  \hfill {\it if {\bf T}={\tt s,c}}   &     in/out       
  & Workspace for the reduced eigenvalue problem         \\
            &  Type(Z){\tt (M0,M0)}  \hfill {\it if {\bf T}={\tt d,z}}    & &  \\ \hline 
{\tt Emin}  &   Type(S)  \hfill {\it if {\bf T}={\tt s,c}}   &  in  & Lower bound of search interval\\   
            &   Type(D)  \hfill {\it if {\bf T}={\tt d,z}}         &    & Hermitian problem \\ \hline
{\tt Emax}  &   Type(S)  \hfill {\it if {\bf T}={\tt s,c}}   &  in  & Upper bound of search interval\\   
            &   Type(D)  \hfill {\it if {\bf T}={\tt d,z}}         &    & Hermitian problem \\ \hline
{\tt Emid}  &  Type(C) \hfill {\it if {\bf T}={\tt s,c}}   &  in    &  Coordinate center of the contour ellipse   \\ 
      & Type(Z) \hfill {\it if {\bf T}={\tt d,z}} & & non-Hermitian problem \\ \hline
{\tt r}  &   Type(S)  \hfill {\it if {\bf T}={\tt d,z}}    &  in  & Horizontal radius of the contour ellipse \\   
            &   Type(D)  \hfill {\it if {\bf T}={\tt s,c}}     &   & non-Hermitian problem  \\ \hline
\end{tabular}
\caption{\label{tab_rci} Definition of arguments specific for the {\tt {\bf T}feast\_{\bf Y}rci\{x\} } interfaces.}
\end{footnotesize}
\end{center}
\end{table}

\subsubsection{RCI Mechanism}


Using the FEAST\_RCI interfaces, the {\tt ijob} parameter must first be  initialized with the value $-1$.
Once the RCI interface is called, the value of the {\tt ijob} output parameter, if different than $0$, 
is used to identify the FEAST operation that needs to be done by the user
Users have then the possibility to customize their own matrix direct or iterative factorization and linear solve 
techniques as well as their own matrix multiplication routine. Table~\ref{tab_ijob} lists all the required cases options needed
using {\tt {\bf \large T}feast\_{\bf  \large Y}rci\{X\}} interfaces, depending on the choices for {\tt \bf {TY}}.
The general reverse communication interface (RCI) mechanism is detailed in Figure~\ref{fig_rci}.


\begin{table}[htbp]
\begin{center}
\begin{footnotesize}
\begin{tabular}{|c|c||c|}
\hline
\tt \bf T & \tt \bf Y & {\tt ijob} parameter values - {\bf cases required}  \\
\hline
\hline
\tt d,s & \tt s & \tt  \{10,11,30,40\}  \\ \hline
\tt z,c & \tt h & \tt  \{10,11,\{20\},21,30,40\} \\ \hline
\tt d,s & \tt g & \tt  \{10,11,\{20\},21,30,31,40,41\}  \\ \hline
\tt z,c & \tt s & \tt    \{10,11,\{20\},21,30,31,40,41\}  \\ \hline
\tt z,c & \tt g & \tt   \{10,11,\{20\},21,30,31,40,41\}   \\ \hline
\hline
\end{tabular}
\caption{\label{tab_ijob} Required options for the {\tt {\bf \large T}feast\_{\bf  \large Y}rci\{x\}} interfaces. 
}
\end{footnotesize}
\end{center}
\end{table}

\begin{figure}[htbp]
\begin{footnotesize}
\begin{lstlisting}[frame=trBL]
ijob=-1 ! initialization
do while (ijob/=0)
 call Tfeast_Yrci{x}({List-rci},fpm,epsout,loop,{List-I},M0,E,X,M,res,info,{Zne,Wne})
 select case(ijob)
 case(10) !!Factorize the complex matrix Az <=(ZeB-A) - or factorize a preconditioner of Az
................ <<< user entry
 case(11) !!Solve the linear system with fpm(23) rhs; Az * Qz=zwork2(1:N,1:fpm(23))
          !!Result in zwork2 <= Qz(1:N,1:fpm(23))
................ <<< user entry
 case(20) !!Factorize (*only if* needed by case(21)) the complex matrix Az'<=Az^H
          !!ATTENTION: The matrix Az from case(10) cannot be overwritten
          !!           - this option would require additional memory storage-  
          !!REMARK:    case(20) becomes obsolete if the solve in case(21) can be performed  
          !!           by reusing the factorization in case(10)
................ <<< user entry
 case(21) !!Solve the linear system with fpm(23) rhs;  Az^H * Qz=zwork2(1:N,1:fpm(23))
          !!Result in zwork2 <= Qz(1:N,1:fpm(23))          
................ <<< user entry
 case(30) !!Perform multiplication A * X(1:N,i:j) result in {z}work1(1:N,i:j)
          !! where i=fpm(24) and j=fpm(24)+fpm(25)-1
................ <<< user entry
 case(31) !!Perform multiplication A^H * X(1:N,i:j) result in {z}work1(1:N,i:j)
          !! where i=fpm(34) and j=fpm(34)+fpm(35)-1
................ <<< user entry
 case(40) !!Perform multiplication B * X(1:N,i:j) result in {z}work1(1:N,i:j)
          !! where i=fpm(24) and j=fpm(24)+fpm(25)-1
................  <<< user entry
 case(41) !!Perform multiplication B^H * X(1:N,i:j) result in {z}work1(1:N,i:j)
          !! where i=fpm(34) and j=fpm(34)+fpm(35)-1
................  <<< user entry
 end select
end do
\end{lstlisting}
\caption{\label{fig_rci} Description of the general FEAST reverse communication interface mechanism (Fortran example).\\
{\bf Remark:} (i) For standard eigenvalue problems {\tt case(40)} and {\tt case(41)} involve only copy operations;
(ii)~If the whole interface is called within an MPI-environment and the code is linked to FEAST\_MPI (i.e. {\tt -lpfeast}), the
operations on the contour integration and the mat-vec operations with multiple rhs, 
will be automatically distributed among the MPI processes.
}
\end{footnotesize}
\end{figure}


\newpage

\subsection{FEAST predefined interfaces}\label{sec_feast_pre}

\subsubsection{Specific declarations}
\noindent For the generalized eigenvalue problem:
\begin{quote}
\begin{tt}
{\large \bf T}feast\_{\large \bf YF}gv~ ({\bf \{List-A\}},{\bf \{List-B\}},fpm,epsout,loop,{\bf \{List-I\}},M0,E,X,M,res,info) \\
{\large \bf T}feast\_{\large \bf YF}gvx ({\bf \{List-A\}},{\{List-B\}},fpm,epsout,loop,{\bf \{List-I\}},M0,E,X,M,res,info,Zne,Wne) 
\end{tt}
\end{quote}
\noindent For the standard eigenvalue problem:
\begin{quote}
\begin{tt}
{\large \bf T}feast\_{\large \bf YF}ev~ ({\bf \{List-A\}},fpm,epsout,loop,{\bf \{List-I\}},M0,E,X,M,res,info) \\
{\large \bf T}feast\_{\large \bf YF}evx ({\bf \{List-A\}},fpm,epsout,loop,{\bf \{List-I\}},M0,E,X,M,res,info,Zne,Wne) 
\end{tt}
\end{quote}
\vspace{0.25cm}
where the series of arguments in each {\tt \bf \{List-A\}}, {\tt \bf \{List-B\}}, and {\tt \bf \{List-I\}}, are specific to the values of {\tt \large \bf T}, {\tt \large \bf Y} and {\large \tt \bf F}, and are given in Table \ref{tab_pre1}. The definition of the arguments 
in {\tt \bf \{List-A\}} and {\tt \bf \{List-B\}} is given in Table \ref{tab_prep}. 
The definitions for the dense, banded and CSR matrix data structures are also provided in the next section.

\begin{table}[htbp]
\begin{center}
\begin{small}
\begin{tabular}{|c|c|c||c|c|c|}
\hline
\tt \bf T & \tt \bf Y & \tt \bf F & {\tt \bf List-A} & {\tt \bf List-B} &{\tt \bf  List-I} \\
\hline
\hline
\multicolumn{3}{|c||}{\tt Dense} & \multicolumn{3}{|c|}{} \\ \hline
\tt z,c & \tt s &\tt  y & \tt  \{ UPLO, N, A, LDA \} &\tt  \{ B, LDB \} &\tt  \{ Emid, r \} \\ \hline
\tt z,c & \tt h & \tt e & \tt \{ UPLO, N, A, LDA \} &\tt  \{ B, LDB \} & \tt \{ Emin, Emax \}\\ \hline
\tt z,c & \tt g &\tt  e & \tt \{ N, A, LDA \} &\tt  \{ B, LDB \} &\tt  \{ Emid, r \}\\ \hline
\tt d,s & \tt s &\tt  y &\tt  \{ UPLO, N, A, LDA \} &\tt  \{ B, LDB \} &\tt  \{ Emin, Emax \}\\ \hline
\tt d,s & \tt g &\tt  e &\tt  \{ N, A, LDA \} &\tt  \{ B, LDB \} &\tt  \{ Emid, r \}\\ \hline
\multicolumn{3}{|c||}{\tt Banded} & \multicolumn{3}{|c|}{} \\ \hline
\tt z,c & \tt s & \tt b &\tt  \{ UPLO, N, kla, A, LDA \} &\tt  \{ klb, B, LDB \} &\tt  \{ Emid, r \}\\ \hline
\tt z,c & \tt h &\tt  b &\tt    \{ UPLO, N, kla, A, LDA \} &\tt  \{ klb, B, LDB \} &\tt  \{ Emin, Emax \} \\ \hline
\tt z,c & \tt g &\tt  b &\tt  \{ N, kla, kua, A, LDA \} &\tt  \{ klb, kub, B, LDB \} &\tt  \{ Emid, r \} \\ \hline
\tt d,s & \tt s &\tt  b &\tt   \{ UPLO, N, kla, A, LDA \} &\tt  \{ klb, B, LDB \} &\tt  \{ Emin, Emax \} \\ \hline
\tt d,s & \tt g &\tt  b &\tt  \{ N, kla, A, LDA \} &\tt  \{ klb, kub, B, LDB \} &\tt  \{ Emid, r \} \\ \hline
\multicolumn{3}{|c||}{\tt Sparse} & \multicolumn{3}{|c|}{} \\ \hline
\tt z,c & \tt s &\tt  csr &\tt  \{ UPLO, N, A, IA, JA \} &\tt  \{ B, IB, JB \} &\tt  \{ Emid, r \} \\ \hline
\tt z,c & \tt h &\tt  csr &\tt  \{ UPLO, N, A, IA, JA \} &\tt  \{ B, IB, JB \} &\tt  \{ Emin, Emax \} \\ \hline
\tt z,c & \tt g &\tt  csr &\tt  \{ N, A, IA, JA \} &\tt  \{ B, IB, JB \} & \tt \{ Emid, r \} \\ \hline
\tt d,s & \tt s &\tt  csr &\tt  \{ UPLO, N, A, IA, JA \} &\tt  \{ B, IB, JB \} &\tt  \{ Emin, Emax \} \\ \hline
\tt d,s & \tt g &\tt  csr &\tt  \{ N, A, IA, JA \} &\tt  \{ B, IB, JB \} &\tt  \{ Emid, r \} \\ \hline
\end{tabular}
\end{small}
\caption{\label{tab_pre1} List of arguments specific for the {\tt {\bf \large T}feast\_{\bf  \large YF}\{ev,gv\}\{x\}} interfaces.} 
\end{center}
\end{table}

\begin{table}[htbp]
\begin{center}
\begin{small}
\begin{tabular}{|l|l|c|l|}
\hline
& & & \\ 
 & Type (Fortran) & Input/ &Description \\
& & Output & \\ 
 &  & & \\ \hline  \hline
{\tt UPLO}  & 
character(len=1) & in & Matrix Storage {\tt ('F','L','U')} \\ 
& & & 'F': Full; 'L': Lower; 'U': Upper\\ \hline
{\tt N}  & integer  & in & Size of the system \\ \hline
{\tt kla}  &    integer          &     in       
  & The number of subdiagonals         \\
   & & &  within the band of {\tt A}.  \\ \hline  
{\tt klu}  &    integer          &     in       
  & The number of superdiagonals         \\
   & & &  within the band of {\tt A}.  \\ \hline  
{\tt klb}  &    integer          &     in       
  & The number of subdiagonals       \\
  & & & within the band of {\tt B}.  \\  
\hline
{\tt kub}  &    integer          &     in       
  & The number of superdiagonals       \\
  & & & within the band of {\tt B}.  \\  
\hline
{\tt A}  &    {\it Same type as {\bf T}}          &     in       
  & Eigenvalue system (Stiffness) matrix         \\
            & {\it with 2D dimension:} ({\tt LDA,N}) {\it if Dense}  & &  \\  
 & {\it with 2D dimension:} ({\tt LDA,N})  {\it if Banded}  & &  \\  
 & {\it with 1D dimension:} ({\tt IA(N+1)-1})  {\it if Sparse}  & &  \\  \hline
{\tt B}  &    {\it Same type as {\bf T}}          &     in       
  & Eigenvalue system (Mass) matrix         \\
   & {\it with 2D dimension:} ({\tt LDB,N})  {\it if Dense}  & &  \\  
 & {\it with 2D dimension:} ({\tt LDB,N})  {\it if Banded}  & &  \\  
 & {\it with 1D dimension:} ({\tt IB(N+1)-1})  {\it if Sparse}  & &  \\  \hline
{\tt LDA}  &    integer          &     in       
  & Leading dimension of {\tt A}; \\
 & & & {\tt LDA>=N}  {\it if Dense} \\   
 & & &  {\tt LDA>=2kla+1}   {\it if Banded; {\tt UPLO='F'}}        \\
& & &  {\tt LDA>=kla+1}   {\it if Banded; {\tt UPLO/='F'}}    \\
\hline  
{\tt LDB}  &    integer          &     in       
  & Leading dimension of {\tt B};  \\
 & & & {\tt LDB>=N}  {\it if Dense} \\   
 & & &  {\tt LDB>=2klb+1}  {\it if Banded; {\tt UPLO='F'}}        \\
& & &  {\tt LDB>=klb+1}  {\it if Banded; {\tt UPLO/='F'}}    \\
\hline
{\tt IA}  &   integer(N+1)          &     in       
  & Sparse CSR Row array of {\tt A}.          \\ \hline
{\tt JA}  &   integer(IA(N+1)-1)          &     in       
  & Sparse CSR Column array of {\tt A}.          \\ \hline
{\tt IB}  &   integer(N+1)          &     in       
  & Sparse CSR Row array of {\tt B}.          \\ \hline
{\tt JB}  &   integer(IB(N+1)-1)          &     in       
  & Sparse CSR Column array of {\tt B}.          \\ \hline
\end{tabular}
\end{small}
\caption{\label{tab_prep} Definition of arguments specific for the 
{\tt {\bf \large T}feast\_{\bf  \large YF}\{ev,gv\}\{x\}} interfaces.}
\end{center}
\end{table}

~
\newpage

\subsubsection{Matrix storage}\label{sec_format}

Let us consider a standard eigenvalue problem and the following (stiffness) matrix {\bf A}: \\[5pt]

\begin{equation}
{\bf A}=
\begin{pmatrix}
 a_{11} & a_{12} & 0 & 0\\
a_{21} & a_{22} & a_{23}& 0  \\
0 & a_{32} & a_{33} & a_{34} \\
0 & 0 & a_{43} & a_{44} \\  
 \end{pmatrix}
\end{equation}\\[5pt]
where $a_{ij}=a_{ji}^*$ for $i\neq j$ (i.e. $a_{ij}=a_{ji}$ if the matrix is real).
Using the FEAST predefined interfaces, this matrix could be stored in dense, banded or sparse format as follows:
\begin{itemize}

\item Using the dense format, {\bf A} is stored in a two dimensional array  in a straightforward fashion.
Using the options {\tt UPLO='L'} or {\tt UPLO='U'}, the lower triangular and upper triangular part respectively,
do not need to be referenced.

\item Using the banded format, {\bf A} is also stored in a two dimensional array  following the 
banded LAPACK-type storage:\\[5pt]
$$
{\bf A=}
\begin{pmatrix}
* & a_{12} & a_{23}& a_{34}  \\
a_{11} & a_{22} & a_{33}& a_{44}  \\
a_{21} & a_{32} & a_{43} & * 
 \end{pmatrix}
$$\\[5pt]
In contrast to LAPACK, no extra-storage space is necessary since {\tt LDA>=2*kla+1} if {\tt UPLO='F'} (LAPACK banded storage 
would require {\tt LDA>=3*kla+1}). For this example, the number of subdiagonals or superdiagonals is {\tt kla=1}. 
Using the option  {\tt UPLO='L'} or {\tt UPLO='U'}, the {\tt kla} rows respectively 
above or below the diagonal elements row, do not need to be referenced (or stored).

\item Using the sparse storage, the non-zero elements of {\bf A} are stored using a set of one 
dimensional arrays ({\tt A,IA,JA})  following the definition of the CSR (Compressed Sparse Row) format
$$
\begin{array}{rl}
{\bf A=}&(a_{11},a_{12},a_{21},a_{22},a_{23},a_{32},a_{33},a_{34},a_{43},a_{44}) \\ 
{\bf IA=}&(1,3,6,9,11) \\
{\bf JA=}&(1,2,1,2,3,2,3,4,3,4) 
\end{array}
$$
Using the option {\tt UPLO='L'} or {\tt UPLO='U'}, one would get respectively

$$
\begin{array}{rl}
{\bf A=}&(a_{11},a_{21},a_{22},a_{32},a_{33},a_{43},a_{44}) \\ 
{\bf IA=}&(1,2,4,6,8) \\
{\bf JA=}&(1,1,2,2,3,3,4) 
\end{array}
\mbox{~~~~and~~~~} \hfill
\begin{array}{rl}
{\bf A=}&(a_{11},a_{12},a_{22},a_{23},a_{33},a_{34},a_{44}) \\ 
{\bf IA=}&(1,3,5,7,8) \\
{\bf JA=}&(1,2,2,3,3,4,4) 
\end{array}
$$

\end{itemize}

Finally, the (mass) matrix {\bf B} that appears in generalized eigenvalue systems, 
should use the same family of storage format than the matrix {\bf A}. It should be noted, however, 
that the bandwidth can be different for the banded format ({\tt klb} can be different than {\tt kla}),
 and the position of the non-zero elements can also be different for the sparse format (CSR coordinates {\tt IB,JB} can
be different than  {\tt IA,JA}).

\newpage

\section{FEAST Parameters and Search Contour}

\subsection{Input FEAST parameters}
 
In the common argument list, the input parameters for the FEAST algorithm are contained into an integer array
of size 64 named here {\tt fpm}. Prior calling the FEAST interfaces, this array needs to be initialized
using the routine {\tt feastinit} as follows (Fortran notation):
\begin{quote}
{\tt call feastinit(fpm)}
\end{quote}

All input FEAST parameters are then set to their default values. The detailed list of these parameters is given in
Table  \ref{tab_fpm}.


\begin{table}[htbp]
\begin{center}
\begin{footnotesize}
\begin{tabular}{|c|l|c|}
\hline
& &  \\ 
{\tt fpm(i)} Fortran & Description & Default value \\ 
{\tt fpm[i-1]} C &  &  \\ \hline  \hline
{\tt i=1}  & 
Print runtime comments on screen  (0: No; 1: Yes) & 0 \\  \hline
{\tt i=2}  & \# of contour points for Hermitian FEAST (half-contour)  & 8 \\ 
  &  if {\tt fpm(16)=0,2}, values permitted (1 to 20, 24, 32, 40, 48, 56)  &  \\ 
&  if {\tt fpm(16)=1}, all values permitted  &  \\ \hline
{\tt i=3}  &  Stopping convergence criteria for double precision ($\epsilon=10^{\tt -fpm(3)}$)  & 12 \\ \hline
{\tt i=4}  &  Maximum number of FEAST refinement loop allowed ($\geq 0$)  & 20 \\ \hline
{\tt i=5}  &  Provide initial guess subspace (0: No; 1: Yes) & 0 \\ \hline
{\tt i=6}  &  Convergence criteria (for the eigenpairs in the search interval) & 1  \\ 
& 0: Using relative error on the trace {\tt epsout} i.e. {\tt epsout$< \epsilon$}   &     \\ 
& 1: Using relative residual {\tt res} i.e. $\max_i {\tt res(i)}< \epsilon$   &     \\ \hline
{\tt i=7}  &  Stopping convergence criteria for single precision ($\epsilon=10^{\tt -fpm(7)}$)  & 5 \\ \hline
{\tt i=8}  & \# of contour points for non-Hermitian FEAST (full-contour)  & 16 \\ 
  &  if {\tt fpm(17)=0}, values permitted (2 to 40, 48, 64, 80, 96, 112)  &  \\ 
&  if {\tt fpm(17)=1}, all values permitted ($>$2)  &  \\ \hline
{\tt i=9}  &  User defined MPI communicator for a given search interval & {\footnotesize MPI\_COMM\_WORLD} \\ \hline
{\tt i=10}  &  Store factorizations with the predefined interfaces (0: No; 1: Yes).  & 0 \\ \hline
{\tt i=14} &  1: FEAST normal execution; 1: Return subspace Q after 1 contour; & 0 \\ 
           &  2: Estimate \#eigenvalues inside search interval  &  \\ \hline
{\tt i=16} &   Integration type for Hermitian (0: Gauss; 1: Trapezoidal; 2: Zolotarev) & 0 \\ \hline
{\tt i=17} &   Integration type for non-Hermitian (0: Gauss, 1: Trapezoidal) & 1 \\ \hline
{\tt i=18} &   Ellipse contour ratio - {\tt fpm(18)}/100 = ratio  'vertical axis'/'horizontal axis'& 100 \\ \hline
{\tt i=19} &   Rotation angle in degree [-180:180] for ellipse using non-Hermitian FEAST & 0 \\
 &   Origin of the rotation is the vertical axis.&  \\ \hline
{\tt i=40-63} & unused & \\ \hline
{\tt All Others}  & Reserved value & N/A \\ \hline
\end{tabular}
\caption{\label{tab_fpm}
List of input FEAST parameters and default values obtained with the routine {\tt feastinit}.\newline
{\bf Remark:} Using the {\tt C} language, the components of the {\tt fpm} array starts at 0 and stops at 63.
Therefore, the components {\tt fpm[j]} in C ({\tt j=0-63}) must correspond to the components {\tt fpm(i)}
in Fortran ({\tt i=1-64}) specified above (i.e. {\tt fpm[i-1]}={\tt fpm(i)}).}
\end{footnotesize}
\end{center}
\end{table}

\newpage
\subsection{Defining a search contour}

\begin{figure}[htbp]
\begin{small}
 \center{\includegraphics[width=0.8\textwidth]
        {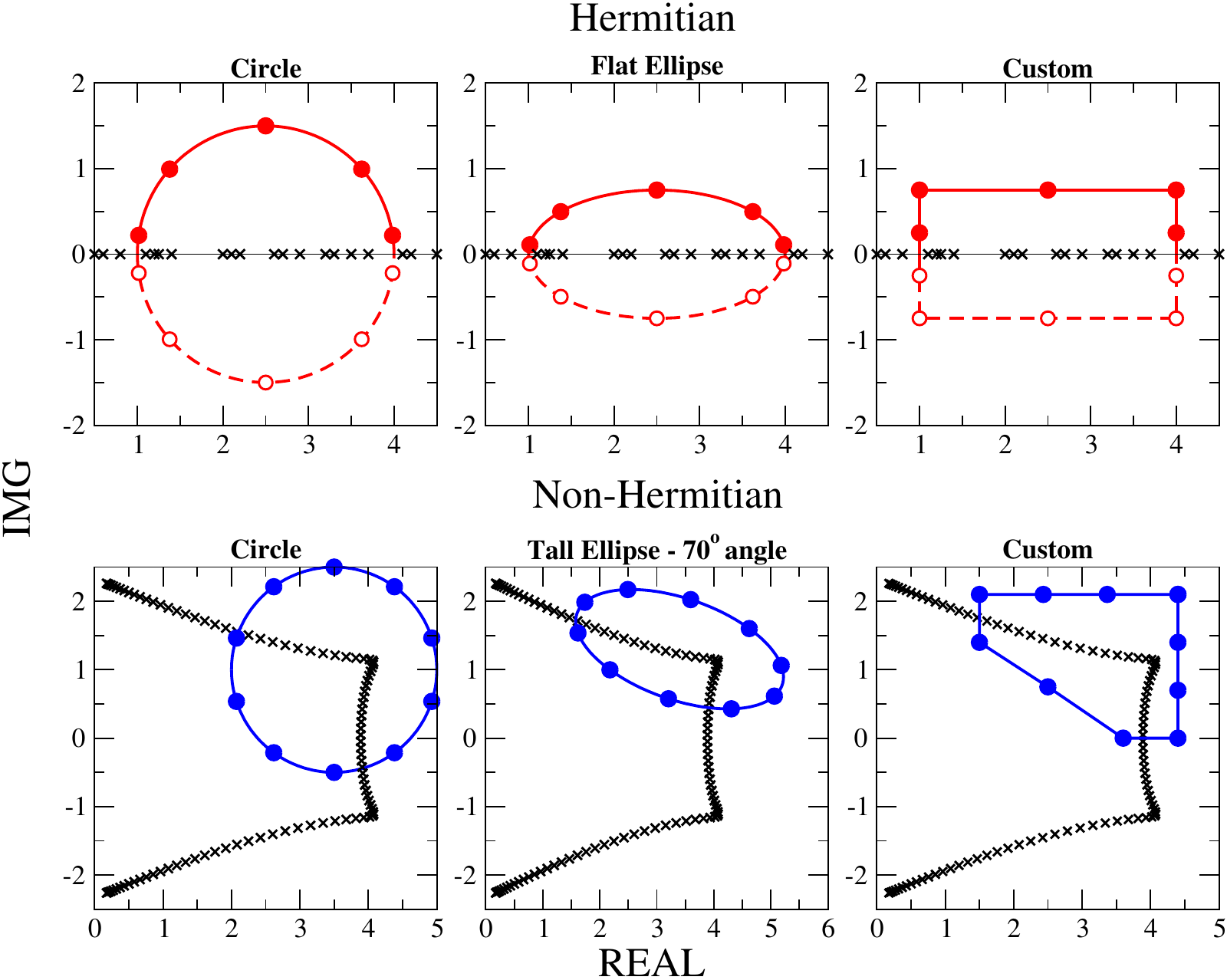}}
        \caption{\label{fig_contour} Various search contour examples for the Hermitian and the non-Hermitian FEAST algorithms.
Both algorithms feature standard ellipsoid contour options and the possibility to define custom arbitrary shapes. 
In the Hermitian case, the contour is symmetric with the real axis and only the nodes in the upper-half may be generated.
In the non-Hermitian case, a full contour is needed to enclose the wanted complex eigenvalues. Some data used to generate these plots:\\
{\bf Hermitian case:} {\tt fpm(2)=5} for all, $[E_{min}, E_{max}]=[1,4]$, $r=1.5$ for all;   {\tt fpm(18)=50} for the flat ellipse; expert routine for
the custom contour \\
{\bf Non-Hermitian case:} {\tt fpm(8)=10} for all; $E_{mid}=3.5+i$ and $r=1.5$ for circle; $E_{mid}=3.4+1.3i$, $r=0.75$,
 {\tt fpm(18)=200}, {\tt fpm(19)=70} for tall rotated ellipse; expert routine for
the custom contour 
}
\end{small}
\end{figure}

Figure \ref{fig_contour} summarizes the different search contour options possible 
for both the Hermitian and non-Hermitian FEAST algorithms.

For the Hermitian case, the user must then specify a 1-dimensional real-valued search interval $[E_{min}, E_{max}]$. 
These two points are used to define a circular or ellipsoid contour $\cal C$ centered on the real axis, and along
 which the complex integration nodes are generated. 
The choice of a particular quadrature rule 
will lead to a different set of relative positions for the nodes and associated quadrature weights. 
Since the eigenvalues are real, it is convenient to select a symmetric contour 
with the real axis (${\cal C}= {\cal C}^*$) since it only requires to operate the quadrature
on the half-contour (e.g. upper half). 

With a non-Hermitian problem, it is necessary to specify  a 2-dimensional search interval that surrounds
the wanted complex eigenvalues. 
Circular or ellipsoid contours can also be used and they can be generated using standard options included into FEAST v3.0. 
These are defined by a complex midpoint $E_{mid}$ and a radius $r$ for a circle (for an ellipse the ratio 
between the horizontal axis $2r$ and vertical axis can also be specified, as well as an angle of rotation). 
in some applications where the eigenvalues of interest belong to a particular subset in the complex plane,  
A ``Custom Contour'' feature is also supported in FEAST v3.0 that  allows to account for 
arbitrary  quadrature nodes and weights.

\newpage
\subsection{Output FEAST {\tt info} details}

Errors and warnings encountered during a run of the FEAST package
are stored in an integer variable, {\tt info}. If the value of the output {\tt info} parameter
is different than ``0'', either an error or warning  was encountered.
The possible return values for the {\tt info} parameter along with the error code descriptions, 
are given in Table~\ref{tab_info}.

\begin{table}[htbp]
\begin{small}
\begin{center}
\begin{tabular}{|l l p{4.2in}|}
\hline
{\tt info} & Classification & Description \\\hline
$202$ & Error & Problem with size of the system {\tt N} \\ 
$201$ & Error & Problem with size of subspace {\tt M0} \\ 
$200$ & Error & Problem with Emin,Emax or Emid, r \\ 
$(100+i)$ & Error & Problem with $i^{th}$ value of the input FEAST parameter (i.e {\tt fpm(i)}) \\
$6$    & Warning &  FEAST converges but subspace is not bi-orthonormal \\
$5$   & Warning & Only stochastic estimation of \#eigenvalues returned {\tt fpm(14)=2} \\
$4$    & Warning & Only the subspace has been returned using {\tt fpm(14)=1} \\
$3$    & Warning & Size of the subspace {\tt M0} is too small ({\tt M0<=M}) \\
$2$    & Warning & No Convergence (\#iteration loops$>${\tt fpm(4)})\\
$1$    & Warning & No Eigenvalue found in the search interval    \\ \hline
$0$    &  Successful exit& \\ \hline
$-1$ & Error & Internal error for allocation memory \\
$-2$ &  Error    & Internal error of the inner system solver in FEAST predefined interfaces \\
$-3$ &  Error    & Internal error of the reduced eigenvalue solver \\
     &           & {\it Possible cause for Hermitian problem: matrix {\bf B} may not be positive definite} \\
$-(100+i)$    & Error & Problem with the $i^{th}$ argument of the FEAST interface\\
\hline
\end{tabular}
\caption{Return code descriptions for the parameter {\tt info}.\newline
{\bf Remark:} In some extreme cases the return value {\tt info=1} may indicate that FEAST has failed
to find the eigenvalues in the search interval. This situation would appear only if a very large search interval 
is used to locate a small and isolated cluster of eigenvalues (i.e. in case the dimension of the search interval is many 
 orders of magnitude off-scaling). For this case, it is then either recommended to increase the number of contour points 
{\tt fpm(2)} or simply rescale more appropriately the search interval.}
\label{tab_info}
\end{center}
\end{small}
\end{table}

%% file: feast-use.tex
\newpage 

\section{FEAST: General use}

This section briefly presents to the FEAST users a list of specifications 
(i.e. what is needed from users), expectations (i.e. what users should expect from FEAST), 
and directions for achieving performances (i.e. including parallel 
scalability and current limitations).

\subsection{Single search interval and FEAST-SMP}

\noindent {\bf Specifications:} 
\begin{itemize}

\item the search interval and the size of the subspace $\rm M_0$ (overestimation of the number of eigenvalues $\rm M$  within); 
If needed, once a search interval is defined, 
the user can take advantage of fast stochastic estimates for $\rm M$ presented 
in Section \ref{sec:se} (tools for FEAST). 
\item the system matrix in dense, banded or sparse-CSR format if FEAST predefined interfaces are used, or a high-performance complex direct
or iterative system solver and matrix-vector multiplication routine if FEAST RCI interfaces are used instead.
\end{itemize}

\noindent {\bf Expectations:}  
\begin{itemize}
\item robust and systematic convergence  to very high accuracy seeking up to $1000$'s  eigenpairs 
(no known failed case for the Hermitian problem);
\item the convergence rate depends on a trade-off between the choice of the search subspace size $\rm M0$ and
the number of contour points (and nature of the quadrature)\footnote{P. Tang, E. Polizzi, SIMAX  35(2), 354–390 - (2014)}. 
For most applications FEAST Hermitian 
will convergence to machine precision in 3 iterations, using  a Gauss-Legendre quadrature with ${\rm M_0 \geq 1.5M}$ and  ${\tt fpm(2) = 8}$. 
If the convergence is too slow, you can: (i) keep on increasing $\rm M_0$ or/and {\tt fpm(2)} ({\tt fpm(8)} for non-Hermitian) 
for the Gauss-Legendre or Trapezoidal quadrature; (ii) decrease the ellipse ratio 
{\tt fpm(18)} for the Hermitian problem; (iii) choose a more robust approach consisting
 of using the Zolotarev quadrature for the Hermitian problem with {\tt fpm(16)=2} 
 (the subspace does not need to be large $\rm M_0\sim M$).

\end{itemize}

\noindent {\bf Directions for achieving performances:} 
\begin{itemize}
\item  $\rm M_0$ should be much smaller than the size of the eigenvalue problem, then the arithmetic complexity 
should mainly depend on the inner system solver (i.e. $\rm O(NM_0)$ for narrow banded or sparse system);
\item storing the factorizations with option {\tt fpm(10)=1} for the predefined interfaces,
 will significantly improve the performances (for FEAST DENSE in particular), 
but can significantly increase the memory usage ($\sim\times${\tt {fpm(2)}} or {\tt {fpm(8)}} for non-Hermitian);   
\item parallel scalability performances at the third level of parallelism depends on the shared memory capabilities of the inner system 
solver i.e. via the shell variable {\tt MKL\_NUM\_THREADS} if a Intel-MKL solver is used (LAPACK or MKL-PARDISO) or the
the shell variable {\tt OMP\_NUM\_THREADS} if SPIKE-SMP is used for the banded interfaces;
\item if $\rm M_0$ increases significantly for a given search interval, 
the complexity $\rm O(M_0^3)$ for solving the reduced system could 
become  significant (typically, if  $\rm M_0>2000$).
In this case it is recommended to consider multiple search intervals to be solved in parallel.
For example, if $10^4$ eigenpairs of a very large system are needed, many
search intervals could be used simultaneously to decrease the size of the reduced dense generalized
eigenproblem (e.g. if $10$ intervals are selected the size of each reduced problem would then be $\sim 10^3$);
\item For very large general sparse and challenging systems, it is strongly recommended for expert application users to make use of
 FEAST-RCI with customized highly-efficient system solvers such as: domain decompositions, 
or iterative solvers with/without preconditioners;  
\end{itemize}

\subsection{Single search interval and FEAST-MPI}\label{sec:mpi}

\noindent {\bf Specifications:} 
\begin{itemize}
\item Same general specification than for the FEAST-SMP; 
\item MPI environment application code and link to the FEAST-MPI library.
\end{itemize}

\noindent {\bf Expectations:}  
\begin{itemize}
\item same general expectation than for FEAST-SMP;
\item ideally, linear scalability performances with the number of MPI processes 
up to the number of linear systems (i.e. Factorization stage) $N_e$ to perform by contour.
If the number of MPI processes is exactly (optimally) equal to $N_e$ which is equal to
either {\tt fpm(2)} for FEAST Hermitian or {\tt fpm(8)} for FEAST non-Hermitian, 
the factorization of the system matrices is  kept in memory 
along the FEAST iterations and ``superlinear scalability'' can then be expected. 
We note one particular case: if the system matrix is real non-symmetric with {\tt Img\{Emid\}=0}, and 
 with {\tt fpm(8)} even number, the number of factorizations becomes {\tt fpm(8)}/2.


 \end{itemize}


\noindent {\bf Directions for achieving performances:} 
\begin{itemize}
\item same general directions than for FEAST-SMP;
\item for a given search interval, the second and third level of parallelism is then easily achieved by MPI (along the contour points) 
calling OpenMP (i.e shared memory linear system solver). Among the two levels of parallelism offered here, there is a trade-off between
the choice of the number of MPI processes and threads by MPI process.  For example let us consider: (i) the use of FEAST-SPARSE interface, (ii)
$8$ contour points (i.e. $8$ linear systems to solve), and (iii) a cluster of $4$ physical nodes with $8$ cores/threads by nodes; 
the two following options (among many others) should be considered to run the MPI code {\tt myprog}:

 $${\tt>mpirun -genv~ MKL\_NUM\_THREADS~ 8 -ppn~1 -n~4~ ./myprog}$$ 
where  $8$ threads can be used on each node, and where each physical node ends up solving consecutively two linear systems using $8$ threads.
This option saves memory.

 $${\tt>mpirun -genv~ MKL\_NUM\_THREADS~ 4 -ppn~2 -n~8~ ./myprog}$$
where $4$ threads can be used on each node, and where the MPI processes end up solving simultaneously the $8$ linear systems using $4$ threads. 
This option should provide better performance but it is more demanding in memory (two linear systems are stored on each node).

In contrast if $8$ physical nodes are available, the best possible option for $8$ contour points becomes:
 $${\tt>mpirun -genv~ MKL\_NUM\_THREADS~ 8 -ppn~1 -n~8~ ./myprog}$$ 
where all the $64$ cores are then used.

\item If more physical nodes than contour points are available, scalability cannot be achieved at the second level parallelism anymore,
but multiple search intervals could then be considered (i.e. first level of parallelism).

\end{itemize}

\subsection{Multiple search intervals and FEAST-MPI}

\begin{figure}[htbp]
\begin{center}
\begin{footnotesize}
\includegraphics[width=0.77\linewidth,angle=0]{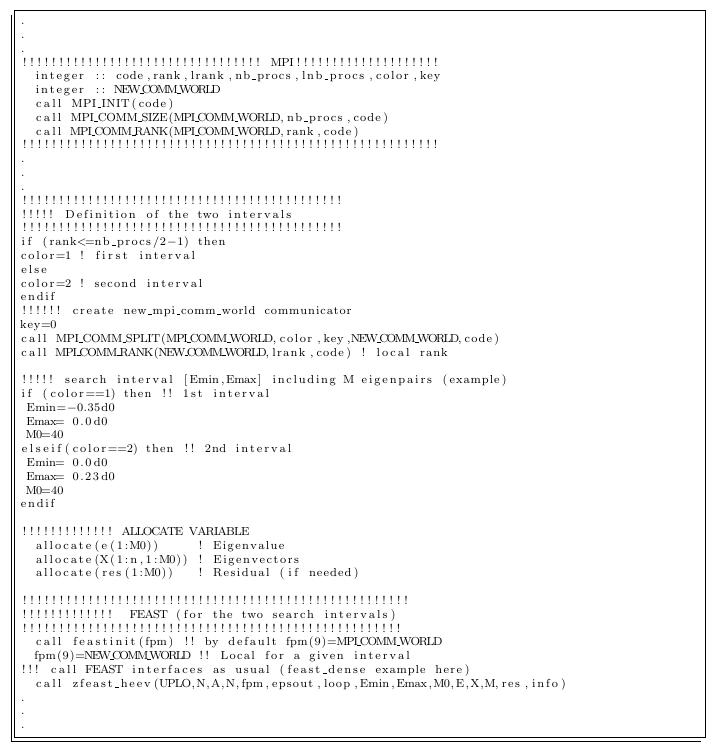}
\caption{\label{3plevels} A simple MPI-F90 example procedure using 
two search intervals and the flag {\tt fpm(9)}. 
In this example, if the code is executed using {\tt n} MPI-processes, the first search interval would be using {\tt n/2} processes 
while {\tt n-n/2} will be used by the second search interval. Driver examples that include three levels of parallelism can be found in 
all subdirectories of {\tt \feastdir/example/Fortran-MPI/} and  {\tt \feastdir/example/C-MPI/} (see FEAST application section for more details).}
\end{footnotesize}
\end{center}
\end{figure}

\noindent {\bf Specifications:} 
\begin{itemize}
\item same general specification than for the FEAST-MPI using a single search interval; 
\item a new flag {\tt fpm(9)} can easily by defined by the user to set up a local MPI communicator
associated to different cluster of nodes for each search interval (which is set to {\tt MPI\_COMM\_WORLD} by default). An
 example on how one can proceed for two search intervals is given in Figure \ref{3plevels}. This can easily be generalized to any numbers
of search intervals.
\end{itemize}

\noindent {\bf Expectations:}  
\begin{itemize}
\item same general expectation than for FEAST-MPI using a single search interval;
\item perfect parallelism using multiple neighboring search intervals without overlap (overall orthogonality should also be preserved).
\end{itemize}

\noindent {\bf Directions for achieving performances:} 
\begin{itemize}
\item same general directions than for FEAST-MPI using a single search interval;
\item in practical applications, the users should have 
a good apriori estimates of the distribution of the overall eigenvalue spectrum in order to make an efficient use of the first level of parallelism 
(i.e. in order to specified the search intervals). Users can currently take advantage
of  fast stochastic estimates presented in Section \ref{sec:se}.
Future developments of FEAST will include  runtime automatic strategies to partition the 
search intervals.
\end{itemize}

%% file: feast_applications.tex
\newpage

\section{FEAST in action}



\subsection{Examples: Hermitian/Non-Hermitian; Fortran/C/MPI; Dense/Banded/Sparse}\label{sec_example}
The {\tt \$FEASTROOT/example} directory provides Fortran, Fortran-MPI, C and MPI-C examples for using the FEAST predefined interfaces.
The Fortran examples are written in F90 but it should be straightforward for the user
 to transform then into F77 if needed (since FEAST uses F77 argument-type interfaces).
Examples include four particular types of eigenvalue problems:
\begin{description}
\item[Example 1] a ``real symmetric'' generalized eigenvalue problem $\bf Ax=\lambda Bx$ , where  
$\bf A$ is real symmetric and $\bf B$ is symmetric positive definite.
$\bf A$ and $\bf B$ are of the size $\rm N=1671$ and
have the same sparsity pattern with number of non-zero elements $\rm NNZ=11435$.  
\item[Example 2] a ``complex Hermitian'' standard eigenvalue problem $\bf Ax=\lambda x$, where  
$\bf A$ is complex Hermitian. 
$\bf A$ is of size $\rm N=600$ with number of non-zero elements $\rm NNZ=2988$.
\item[Example 3] a ``real non-symmetric'' generalized eigenvalue problem $\bf Ax=\lambda Bx$ , where  
$\bf A$ and $\bf B $ are considered real non-symmetric.
$\bf A$ and $\bf B$ are of the size $\rm N=1671$ and
have the same sparsity pattern with number of non-zero elements $\rm NNZ=13011$. 
\item[Example 4] a ``complex symmetric'' standard eigenvalue problem $\bf Ax=\lambda x$, where  
$\bf A$ is complex symmetric. 
$\bf A$ is of size $\rm N=801$ with number of non-zero elements $\rm NNZ=24591$.

\end{description} 

Examples are solved in double precision arithmetic (Example 1 and 2 also include single precision drivers)
and either a dense, banded or sparse storage for the matrices.

The {\tt \$FEASTROOT/example/\{Hermitian,non-Hermitian\}} directories contain the 
subdirectories {\tt Fortran, C, Fortran-MPI, C-MPI} which, in turn, contain similar
subdirectories {\tt 1\_dense}, {\tt 2\_banded}, {\tt 3\_sparse} with source code drivers for the above four examples 
in single and double precisions. 

In order to compile and run the examples of the FEAST package, please follow the following steps:
\begin{enumerate}
\item Go to the directory {\tt \$FEASTROOT/example}
\item Edit the {\tt make.inc} file and follow the directions to customize appropriately: (i) the name/path of your F90, C and MPI compilers
(if you wish to compile and run the F90 examples alone, it is not necessary to specify the C compiler as well as MPI and vice-versa);
 (ii) the path {\tt LOCLIBS} for the FEAST libraries and both paths and names in {\tt LIBS} of the  MKL-PARDISO, LAPACK and BLAS libraries
(if you do not wish to compile the sparse examples there is no need to specify the path and name of the MKL-PARDISO library).

By default, all the options in the {\tt make.inc} assumes calling the FEAST library compiled with no-runtime dependency (add the appropriate
flags {\tt -lifcoremt, -lgfortran, etc.} otherwise). Additionally,  {\tt LIBS} in {\tt make.inc} uses by default 
the  Intel-MKL  (v10.x or later) to link the required libraries.
\item By executing {\tt make},   all the different Makefile options will be printed out including
 compiling alone, compiling and running, and cleaning. For example, 
\begin{quote}
{\tt >make allF}
\end{quote}
compiles all Fortran examples, while
\begin{quote}
{\tt >make rallF}
\end{quote}
compiles and then run all Fortran examples.
\item If you wish 
to compile and/or run a given example, for a particular language and with a particular storage format, 
just go to one the corresponding subdirectories  {\tt 1\_dense}, {\tt 2\_banded}, {\tt 3\_sparse} of
the directories {\tt Fortran, C, Fortran-MPI, C-MPI}. You will find
a local {\tt Makefile} (using the same options defined above in the {\tt make.inc} file) 
as well as the source codes covering the examples above.  The Hermitian 
{\tt 1\_dense} directories include also the ``helloworld'' source code examples
presented in Section \ref{sec_hello}. The Hermitian {\tt Fortran-MPI} and {\tt Fortran-C} directories contain one additional example using 
all three levels of parallelism for FEAST. The non-Hermitian directories contain one additional example
 for generating a custom complex contour for the system matrix in example 4 above.
\end{enumerate}


\subsection{FEAST utility sparse drivers}
If a sparse matrix can be provided by the user in coordinate/matrix market format,
the {\tt \$FEASTROOT/utility} directory offers a quick way  
to test all the FEAST parameter options and 
the efficiency/reliability of the FEAST SPARSE predefined interfaces using the MKL-PARDISO solver. 
Two general drivers are provided for FEAST-SMP and FEAST-MPI,  named {\tt driver\_feast\_sparse}
or  {\tt pdriver\_feast\_sparse} in the respective subdirectories.

You will also find a local {\tt make.inc} file where compiler and libraries paths/names 
need to be edited and changed appropriately (the MKL-PARDISO solver is needed). The command
``{\tt>make all}'' should compile the drivers.

If we denote {\tt mytest} a generic name for the user's eigenvalue system test $\bf Ax=\lambda x$
or  $\bf Ax=\lambda Bx$. You will need to create the following three files:
     \begin{itemize}
       \item {\tt mytest\_A.mtx} should contain the matrix {\bf A} in coordinate format; As a reminder,
 the coordinate format is defined row by row as
\begin{footnotesize}
\begin{lstlisting}[frame=trBL]
      N       N       NNz
      :       :       :              :
      i       j      real(valj)    img(valj)
      :       :       :              : 
      :       :       :              :
      :       :       :              :
     iNNZ    jNNZ    real(valNNZ) img(valNNZ)
\end{lstlisting}
 \end{footnotesize}
with {\tt N}: size of matrix, and {\tt NNZ:} number of non-zero elements.
       \item {\tt mytest\_B.mtx} should contain the matrix {\bf  B} (if any) in coordinate format;
       \item {\tt mytest.in} should contain the search interval, selected FEAST parameters, etc. 
The following {\tt .in} file is given as a template example (here for solving a standard eigenvalue problem in 
double precision):
\begin{footnotesize}
\begin{lstlisting}[frame=trBL]
s       ! s: symmetric, h: hermitian, g: general
g       ! e=standard or g=generalized eigenvalue problem
d       ! (s,d,c,z) precision i.e (single real,double real,complex,double complex)
F       ! UPLO (L: lower, U: upper, F: full) for the coordinate format of matrices
0.18d0  ! Emin
1.00d0  ! Emax
25      ! M0 search subspace (M0>=M)
2       !!!!!!!!!! How many changes from default fpm[1,64] (use 1-64 indexing)
1 1     !fpm(1)[0]=1 !example comments on/off  (0,1)
10 1    !fpm(10) - factorizations saving- on/off (0,1)
\end{lstlisting}   
\end{footnotesize}
You may change any of the above options to fit your needs. For example, you could add as many fpm FEAST parameters as you wish.
It should be noted that the {\tt UPLO} {\tt L} or {\tt U} options give you the possibility
to provide only the lower or upper triangular part of the matrices {\tt mytest\_A.mtx} and {\tt mytest\_B.mtx} 
 in coordinate format.
\end{itemize}

Finally results and timing can be obtained by running the FEAST-SMP sparse driver: 
\begin{quote}
{\tt >\$FEASTROOT/utility/Fortran/driver\_feast\_sparse <PATH\_TO\_MYTEST>/mytest} 
\end{quote}            

For the 
FEAST-MPI sparse drivers, a run would look like (other options could be applied including Ethernet fabric, etc.):

 {\tt mpirun  -genv~ MKL\_NUM\_THREADS~ <y> -ppn~ 1 -n <x> \BS \\   
      \Indent  \$FEASTROOT/utility/Fortran/pdriver\_feast\_sparse  <PATH\_TO\_MYTEST>/mytest} \\

where ${\tt <x>}$ represents the number of nodes at the second level of parallelism (along the contour).
As a reminder, the third level of parallelism for MKL-PARDISO (for the linear system solver) is activated  by the setting shell variable
 {\tt MKL\_NUM\_THREADS} equal to the desired number of threads. In the example above for MPI, if the  {\tt MKL\_NUM\_THREADS} is set with value 
{\tt <y>}, i.e. FEAST would run on {\tt <x>} separate nodes, each one using {\tt <y>} threads. Several  combinations of {\tt <x>} and {\tt <y>} are
possible depending also on the value of the {\tt -ppn} directive.\\

In order to illustrate a direct use of the utility drivers,  several examples are provided 
in the directory {\tt \$FEASTROOT/utility/data} summarized in Table~\ref{tab_data}.

\begin{table}
\begin{small}
\begin{center}
\begin{tabular}{|l|cc||ccc||cc|}
\hline
                &  Real   &  Complex  &  Symmetric &  Hermitian &  General &  Standard &  Generalized \\ \hline
helloworld      &    X   &            &        X   &            &           &         X  &            \\ \hline
system1         &  X     &            &      X     &            &           &            &       X      \\ \hline
system2         &        &     X      &            &       X    &           &       X    &             \\ \hline
system3         &  X     &            &            &            &       X   &            &       X     \\ \hline
system4         &        &     X      &      X     &            &           &       X    &             \\ \hline
cnt             &  X     &            &      X     &            &           &            &       X     \\ \hline
co              &  X     &            &      X     &            &           &            &       X     \\ \hline
c6h6            &  X     &            &      X     &            &           &            &       X      \\ \hline
Na5             &  X     &            &      X     &            &           &       X    &             \\ \hline
grcar           &  X     &            &            &            &       X   &       X    &             \\ \hline
qc324           &        &     X      &      X     &            &           &       X    &              \\ \hline
bcsstk11        &  X     &            &      X     &            &           &            &       X       \\ \hline
\end{tabular}
\end{center}
\caption{\label{tab_data} List of system matrices provided in the {\tt \$FEASTROOT/utility/data} directory. System 1 to 4 corresponds
to the matrices used in the {\tt example} directory.}
\end{small}
\end{table}

To run a specific test, you can execute (using the Fortran driver for example): 
\begin{quote}
{\tt >\$FEASTROOT/utility/Fortran/driver\_feast\_sparse ../data/cnt} 
\end{quote}

%% file: feast_tools.tex
\newpage

\section{Additional Tools for FEAST}

\subsection{Stochastic estimate}\label{sec:se}
To make use of FEAST, the user must (i) select a search interval or contour in the complex plane, (ii) provide a search subspace size
that should overestimate the number of eigenvalue {\tt M} that are inside the search contour. FEAST provides options to obtain
a stochastic estimate for {\tt M} \footnote{{\it Efficient Estimation of Eigenvalue Counts in an Interval,} \\E. Di Napoli , E. Polizzi, Y. Saad,\url{http://arxiv.org/abs/1308.4275}}.
 If the flag {\tt fpm(14)} is set to 2, FEAST will perform a single contour and return
the estimated value of {\tt M}. In turn, the value in {\tt res(1:M0)} will return the running average of the estimation 
(i.e. {\tt M$\equiv$ res(M0)}). For this FEAST run, a good estimate can be obtained even if the search subspace size {\tt M0} is chosen rather
 small $\sim 10$. In addition, the whole operation does not require a lot of accuracy to succeed, for example: (i) the flag {\tt fpm(2)}
(i.e. number of linear systems to solve) could be set to $\sim 3$; (ii) if iterative solvers such as GMRES are used 
in the RCI interface, the stopping criteria could be chosen very modest ($\sim 10^{-2}$); (iii) 
the single precision routines could be used as well.

\subsection{Custom contours}

Only eigenvalues inside of the user defined interval are calculated with FEAST.  Since eigenvalues of Hermitian (and real-symmetric) matrices are real this interval can be defined by [$\lambda _{min}$, $\lambda _{max}$]. Non-Hermitian matrices possess a complex eigenspectrum and require a 2-dimensional interval. The Custom Contour feature grants the flexibility to target specific eigenvalues. This feature is available for both Hermitian and non-Hermitian codes and must be used with ``Expert'' routines that take two additional arguments containing the complex integration nodes and weights. Custom contours can be employed by following three simple steps: 
\begin{enumerate}
\itemsep 1pt
\parskip 1pt
\item Define a contour (half-contour that encloses [$\lambda _{min}$, $\lambda _{max}$] for the Hermitian problem, or full
contour for the non-Hermitian problem), 
\item Calculate corresponding integration nodes and weights, and 
\item Call ``Expert'' FEAST routine (either predefined  or RCI interfaces).
\end{enumerate}
FEAST provides a   routine {\tt \{C,Z\}FEAST\_customcontour} that can assist the user to extract 
nodes and weights from a custom design arbitrary geometry in the complex plane.
This routine is only useful for the non-Hermitian FEAST (full-contour).


\subsubsection{Defining a Custom Contour using  {\tt \{C,Z\}FEAST\_customcontour}}

Users must only define the geometry of their contour. 
The contour can be comprised of line segments and half ellipses. 
Two important points to note: {\bf the actual contour will end up being  a polygon} defined by the integration points 
along the path - and - {\bf only convex contours may be used}. A geometry that contains {\tt P} contour parts/pieces 
is defined using three arrays {\tt Zedge}, {\tt Tedge}, and {\tt Nedge}. The interface is defined below and
the description of the arguments list is given in Table~\ref{tab_custom}.

\begin{quote}
\begin{tt}
\{{\bf C,Z}\}FEAST\_customcontour(Nc,P,Nedge,Tedge,Zedge,Zne,Wne)
\end{tt}
\end{quote}

\begin{table}[htbp]
\begin{center}
\begin{footnotesize}
\begin{tabular}{|l|l|c|l|}
\hline
& & & \\ 
 & Type & Input/Output &Description \\ 
 &  & & \\ \hline  \hline
{\tt Nc}  & 
integer  & in & The total number of integration nodes, should be equal to \\
   &      &    & {\tt SUM({\bf Nedge}(1:P))} \\ 
 &   &   & to be used for {\tt fpm(2)} when calling FEAST \\ \hline
{\tt P}  & 
integer  & in & Number of contour parts/pieces that make up the contour\\ \hline
{\tt Zedge} & integer({\tt P}) & in & Complex endpoints of each contour piece \\
         &                     &   & Remark: * endpoints positioned in clockwise direction \\
         &                     &   &~~~~~~~~~~~~~* the $k^{th}$ piece is [{\tt Zedge($k$),Zedge($k+1$)}] \\  
         &                     &   &~~~~~~~~~~~~~* last piece is [{\tt Zedge(P),Zedge(1)}]  \\ \hline
{\tt Tedge} & integer({\tt P}) & in &   The type of each contour piece: \\ 
            &                  &    &   *If {\tt Tedge($k$)}=0,   $k^{th}$ piece is a line \\
            &                  &    &   *If {\tt Tedge($k$)}$>$0, $k^{th}$ piece is a (convex) half-ellipse \\
            &                  &    &   ~~~~~~~~~~~~~~~~~~~~~ with {\tt Tedge(k)}/100 = ratio $a$/$b$ \\
            &                  &    &   ~~~~~~~~~~~~~~~~~~~~~ and $a$ primary radius from the endpoints \\
            &                  &    &   ~~~~~~~~~~~~~~~~~~~~~ Remark: 100 is a half-circle \\ \hline
{\tt Nedge} & integer({\tt P}) & in & \# integration interval to consider for each piece \\
            &                  &    & define the accuracy of the trapezoidal rule by piece for FEAST\\ \hline
{\tt Zne}  &   Type(Z)({\tt Nc})  \hfill {\it if {\bf T}={\tt Z}}    &  out  & Custom integration nodes for FEAST\\   
            &   Type(C)({\tt Nc})  \hfill {\it if {\bf T}={\tt C}}     &   &   \\ \hline
{\tt Wne}  &   Type(Z)({\tt Nc})  \hfill {\it if {\bf T}={\tt Z}}    &  out  & Custom integration weights for FEAST\\   
            &   Type(C)({\tt Nc})  \hfill {\it if {\bf T}={\tt C}}     &   &   \\ \hline
\end{tabular}
\caption{\label{tab_custom} List of arguments for {\tt \{{\bf T}\}FEAST\_customcontour}.}
\end{footnotesize}
\end{center}
\end{table}

\begin{figure}[htbp]
\begin{minipage}{0.5\linewidth}
\centering
\includegraphics[width=\textwidth,angle=0]{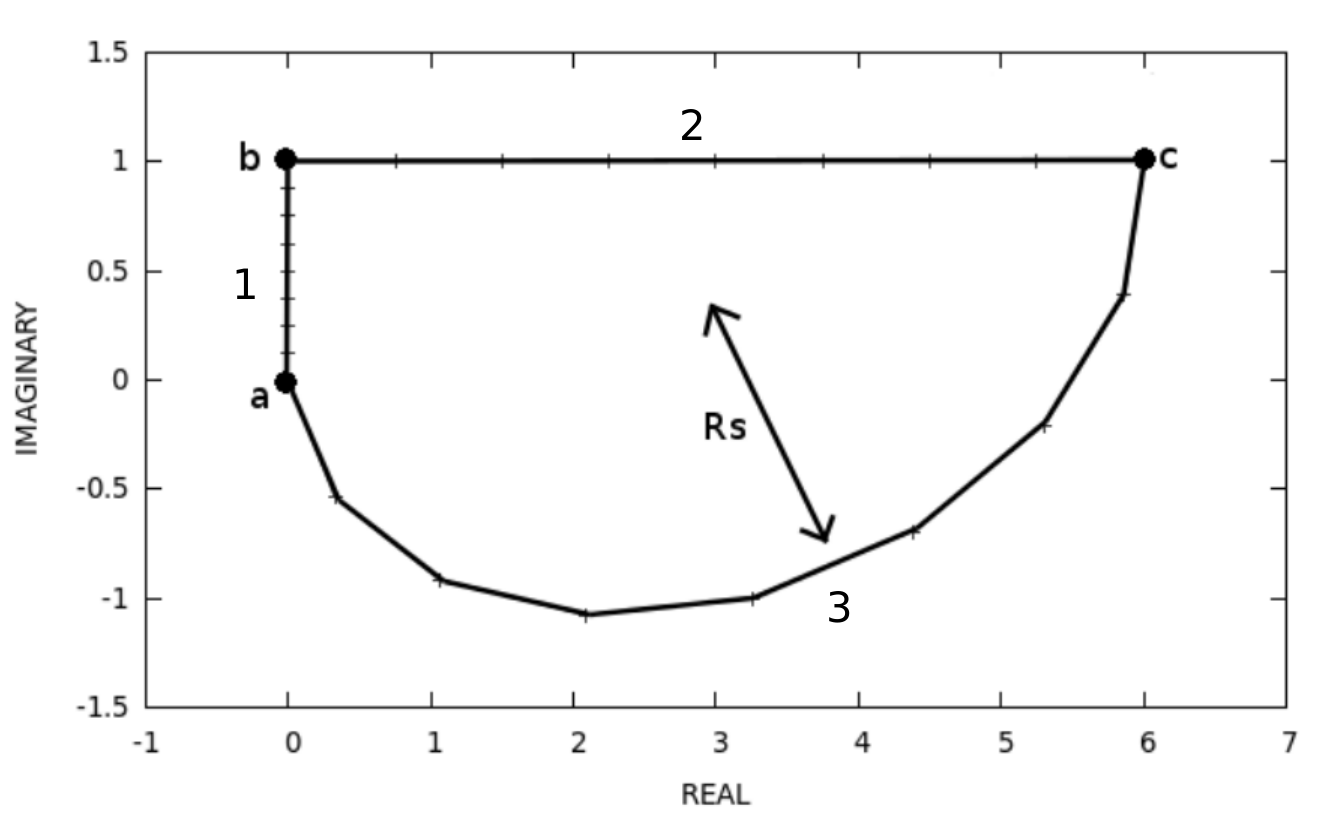}
\end{minipage}\hfill
\begin{minipage}{0.38\linewidth}
\caption{Example contour containing two line segments (pieces 1 and 2) and a half ellipse (piece 3). 
The secondary radius is half of the primary radius defined by nodes a and c. All the {\tt Nc} (24) generated {\tt Zne} points are represented.}
\label{fig_customcontour}
\end{minipage}
\end{figure}

As an example, the following code will generate the complex contour in Figure~\ref{fig_customcontour}.
\noindent
\begin{footnotesize}
\begin{lstlisting} [frame=trBL]
P = 3 ! number of pieces that make up the contour 
allocate( Zedge(1:P),Nedge(1:P),Tedge(1:P) )
Zedge = (/(0.0d0,0.0d0),(0.0d0,1.0d0),(6.0d0,1.0d0)/)
Tedge(:) = (/0,0,50/) ! (line)--(line)--(half-circle)
Nedge(:) = (/8,8,8/)  ! 8 integration intervals for each piece
Nc = sum(Nedge(1:P)) !!User enter # of contour points (here 24)
allocate( Zne(1:Nc), Wne(1:Nc) ) 
call zfeast_customcontour(Nc,P,Nedge,Tedge,Zedge,Zne,Wne)
\end{lstlisting}	
\end{footnotesize}

\subsubsection{Calling the Expert FEAST Routines}

There exists an expert version of each FEAST driver and RCI routine. They are signified by an appending ``X" 
and accept the two additional arguments $Zne$ and $Wne$ containing the integration nodes and weights. 
Below is an example F90 code calling the dense general complex FEAST driver with a custom contour.

\begin{footnotesize}
\begin{lstlisting} [frame=trBL]
program feast_cc
  complex(kind=kind(1.0d0)),dimension(:,:),allocatable :: A, B, X
  complex(kind=kind(1.0d0)),dimension(:),allocatable :: E, Zne, Wne, Zedge
  complex(kind=kind(1.0d0)) :: Emid 
  double precision :: r, epsout
  integer :: M0, P, info, M, N, LDA, LDB
  integer,dimension(64) :: fpm
  double precision,dimension(:),allocatable :: res
  integer,dimension(:),allocatable :: Nedge, Tedge

   allocate(A(1:N,1:N), B(1:N,1:N), E(1:M0), X(1:N,1:2*M0), res(1:2*M0))
   ... Load A,B
   call feastinit(fpm)
   ... Define Custom Contour - Fill up the P elements of Zedge,Tegde,Nedge
   fpm(2) = sum(Nedge(1:P))
   allocate(Zne(1:fpm(2)), Wne(1:fpm(2))) 
   call zfeast_customcontour(fpm(2),P,Nedge,Tedge,Zedge,Zne,Wne) ! generate Zne,Wne   
   call zfeast_gegvx(N,A,LDA,B,LDA,fpm,epsout,loop,Emid,r,M0,E,X,M,res,info,Zne,Wne)
end program feast_cc
\end{lstlisting}	
\end{footnotesize}


\newpage

\subsection{List of all FEAST tool routines}

\hrule	
\vspace{5mm}	
{\Large \bf \texttt - [cz]feast$\_$contour}
\begin{center}
{\texttt call zfeast$\_$contour}($Emin$, $Emax$, $fpm2$, $fpm16$, $fpm18$, $Zne$, $Wne$)
\end{center}

\begin{tabular}{ r  l l p{10cm} }
{\bf Inputs:} & & & \\
$Emin$: & Scalar & Type(S,D) & Lower bound of search interval\\
$Emax$: & Scalar & Type(S,D) & Upper bound of search interval\\
$fpm2$: & Scalar & Type(I) & Number of contour points for integration (half-contour)\\
$fpm16$: & Scalar & Type(I) & Integration type\\
$fpm18$: & Scalar & Type(I) & Ellipse definition\\
{\bf Ouputs:} & & & \\
$Zne$: & Dim($fpm2$) & Type(C,Z) & Integration Nodes (half-contour)\\
$Wne$: & Dim($fpm2$) & Type(C,Z) & Integration Weights (half-contour)\\
\end{tabular}

\begin{itemize}
\item[$\bullet$] Returns FEAST integration nodes and weights for a contour defined by $Emin$ and $Emax$. To be used with complex Hermitian and real symmetric FEAST.
\end{itemize}

\vspace{1mm}
\hrule
\vspace{5mm}
{\noindent \Large \bf \texttt - [cz]feast$\_$gcontour}
\begin{center}
\texttt call zfeast$\_$gcontour( $Emid$ , $r$ , $fpm8$ , $fpm17$ , $fpm18$ , $fpm19$, $Zne$ , $Wne$ )
\end{center}

\begin{tabular}{ r  l l p{10cm} }
{\bf Inputs:} & & & \\
$Emid$: & Scalar & Type(C,Z) & Midpoint of search interval\\
$r$: & Scalar & Type(S,D) & Radius of search interval\\
$fpm8$: & Scalar & Type(I) & Number of contour points for integration (full contour)\\
$fpm17$: & Scalar & Type(I) & Integration type\\
$fpm18$: & Scalar & Type(I) & Ellipse definition\\
$fpm19$: & Scalar & Type(I) & Ellipse rotation angle\\
{\bf Ouputs:} & & & \\
$Zne$: & Dim($fpm8$) & Type(C,Z) & Integration Nodes (full contour)\\
$Wne$: & Dim($fpm8$) & Type(C,Z) & Integration Weights (full contour)\\
\end{tabular}

\begin{itemize}
\item[$\bullet$] Returns FEAST integration nodes and weights for a contour defined by $Emid$ and $r$. To be used with non-Hermitian FEAST (complex-symmetric, real non-symmetric, general-complex).
\end{itemize}

\vspace{1mm}
\hrule
\vspace{5mm}
{\noindent \bf \Large \texttt - [cz]feast$\_$customcontour}
\begin{center}
\texttt call zfeast$\_$customcontour( $fpm8$ , $ccN$ , $Nedge$ , $Tedge$ , $Zedge$ , $Zne$ , $Wne$ )
\end{center}

\begin{tabular}{  r  l l p{8cm}   }
{\bf Inputs:} & & & \\
$fpm8$: & Scalar & Type(I) & Number of contour points: sum($Nedge$(1:$ccN$))\\
$ccN$: & Scalar & Type(I) & Number of segments that comprise contour\\
$Nedge$: & Dim($ccN$) & Type(I) & Number of contour points for each segment\\
$Tedge$: & Dim($ccN$) & Type(I) & Type of each segment\\
$Zedge$: & Dim($ccN$) & Type(C,Z) & Start node of each segment\\
{\bf Outputs:} & & & \\
$Zne$: & Dim($fpm8$) & Type(C,Z) & Integration Nodes\\
$Wne$: & Dim($fpm8$) & Type(C,Z) & Integration Weights\\
\end{tabular}

\begin{itemize}
\item[$\bullet$] Returns FEAST integration nodes and weights for a user defined contour. To be used with FEAST expert routines.
\end{itemize}

\newpage

\vspace{5mm}
\hrule
\vspace{5mm}
{\noindent \Large \bf \texttt - [sd]feast$\_$rational}
\begin{center}
\texttt call dfeast$\_$rational( $Emin$ , $Emax$ , $fpm2$ , $fpm16$ , $fpm18$ , $Eig$ , $M0$ , $f$ )
\end{center}

\begin{tabular}{  r  l l p{8cm}   }
{\bf Inputs:} & & & \\
$Emin$: & Scalar & Type(S,D) & Lower bound of search interval\\
$Emax$: & Scalar & Type(S,D) & Upper bound of search interval\\
$fpm2$: & Scalar & Type(I) & Number of contour points for integration (half-contour)\\
$fpm16$: & Scalar & Type(I) & Integration type\\
$fpm18$: & Scalar & Type(I) & Ellipse definition\\
$Eig$: & Dim($M0$) & Type(S,D) & Set of points to evaluate rational function\\
$M0$: & Scalar & Type(I) & Size of $Eig$/$f$\\
{\bf Outputs:} & & & \\
$f$: & Dim($M0$) & Type(S,D) & Value of rational function at each point in Eig\\
\end{tabular}

\begin{itemize}
\item[$\bullet$] Evaluates rational/selection function for contour defined by $Emin$ and $Emax$ at a set of real values stored in $Eig$.
\end{itemize}
\vspace{5mm}
\hrule
\vspace{5mm}
{\noindent \bf \Large \texttt - [sd]feast$\_$rationalx}

\begin{center}
\texttt call dfeast$\_$rationalx( $Zne$ , $Wne$ ,$fpm2$ ,  $Eig$ , $M0$ , $f$ )
\end{center}

\begin{tabular}{  r  l l p{8cm}   }
{\bf Inputs:} & & & \\
$Zne$: & Dim($fpm2$) & Type(C,Z) & Integration nodes of search interval\\
$Wne$: & Dim($fpm2$) & Type(C,Z) & Integration Weights of search interval\\
$fpm2$: & Scalar & Type(I) & Number of contour points for integration (half-contour)\\
$Eig$: & Dim($M0$) & Type(S,D) & Set of points to evaluate rational function\\
$M0$: & Scalar & Type(I) & Size of $Eig$/$f$\\
{\bf Outputs:} & & & \\
$f$: & Dim($M0$) & Type(S,D) & Value of rational function at each point in Eig\\
\end{tabular}

\begin{itemize}
\item[$\bullet$] Evaluates rational/selection function for contour defined by $Zne$ and $Wne$ at a set of real values stored in $Eig$.
\end{itemize}
\vspace{5mm}
\hrule
\vspace{5mm}
{\noindent \Large \bf \texttt - [cz]feast$\_$grational}

\begin{center}
\texttt call zfeast$\_$grational( $Emid$ , $r$ , $fpm8$ , $fpm17$ , $fpm18$, $fpm19$ , $Eig$ , $M0$ , $f$ )
\end{center}

\begin{tabular}{  r  l l p{8cm}   }
{\bf Inputs:} & & & \\
$Emid$: & Scalar & Type(C,Z) & Midpoint of search interval\\
$r$: & Scalar & Type(S,D) & Radius of search interval\\
$fpm8$: & Scalar & Type(I) & Number of contour points for integration (full-contour)\\
$fpm17$: & Scalar & Type(I) & Integration type\\
$fpm18$: & Scalar & Type(I) & Ellipse definition\\
$fpm19$: & Scalar & Type(I) & Ellipse rotation angle\\
$Eig$: & Dim($M0$) & Type(C,D) & Set of points to evaluate rational function\\
$M0$: & Scalar & Type(I) & Size of $Eig$/$f$\\
{\bf Outputs:} & & & \\
$f$: & Dim($M0$) & Type(C,Z) & Value of rational function at each point in Eig\\
\end{tabular}

\begin{itemize}
\item[$\bullet$] Evaluates rational/selection function for contour defined by $Emid$ and $r$ at a set of complex values stored in $Eig$.
\end{itemize}

\vspace{5mm}
\hrule
\vspace{5mm}
{\noindent \Large \bf \texttt - [cz]feast$\_$grationalx}

\begin{center}
\texttt call zfeast$\_$grationalx( $Zne$ , $Wne$ , $fpm8$ ,  $Eig$ , $M0$ , $f$ )
\end{center}

\begin{tabular}{  r  l l p{8cm}   }
{\bf Inputs:} & & & \\
$Zne$: & Dim($fpm8$) & Type(C,Z) & Integration nodes of search interval\\
$Wne$: & Dim($fpm8$) & Type(C,Z) & Integration Weights of search interval\\
$fpm8$: & Scalar & Type(I) & Number of contour points for integration (full contour)\\
$Eig$: & Dim($M0$) & Type(C,Z) & Set of points to evaluate rational function\\
$M0$: & Scalar & Type(I) & Size of $Eig$/$f$\\
{\bf Outputs:} & & & \\
$f$: & Dim($M0$) & Type(C,Z) & Value of rational function at each point in Eig\\
\end{tabular}

\begin{itemize}
\item[$\bullet$] Evaluates rational/selection function for contour defined by $Zne$ and $Wne$ at a set of complex values stored in $Eig$.
\end{itemize}

\hrule